\documentclass[a4paper,10pt,twocolumn,twoside,journal]{article}

\usepackage{geometry}
\advance\oddsidemargin by -1in
\advance\evensidemargin by -1in
\advance\headsep by 2.8mm
\usepackage[numbers]{natbib}

\usepackage{authblk}
\usepackage{amssymb}
\usepackage{amsmath}
\usepackage{algorithm}
\usepackage{algorithmic}
\usepackage{adjustbox}
\usepackage{hyperref}
\usepackage{subcaption}
\usepackage{array}
\usepackage{url}
\usepackage{tikz}
\usetikzlibrary{quantikz2}

\usepackage{orcidlink}

\def\BibTeX{{\rm B\kern-.05em{\sc i\kern-.025em b}\kern-.08em
    T\kern-.1667em\lower.7ex\hbox{E}\kern-.125emX}}

\usepackage[normalem]{ulem}

\usepackage[compact]{titlesec}

\title{Advanced Scheduling Strategies for Distributed Quantum Computing Jobs}

\author[1]{Gongyu~Ni\orcidlink{0009-0000-0733-4474}}
\author[2]{Davide~Ferrari\orcidlink{0000-0002-4777-7234}}
\author[1]{Lester~Ho\orcidlink{0000-0001-9296-8681}}
\author[2,*]{Michele~Amoretti\orcidlink{0000-0002-6046-1904}}

\affil[1]{\small \textit{Wireless Communications Laboratory}, Tyndall National Institute, Dublin, Ireland}
\affil[2]{\small \textit{Quantum Software Laboratory}, Department of Engineering and Architecture, University of Parma, Parma, 43124 Italy (\href{https://www.qslab.unipr.it/}{https://www.qslab.unipr.it/})}
\affil[*]{\small Corresponding author: Michele Amoretti, michele.amoretti@unipr.it.}

\date{}

\begin{document}

\maketitle

\begin{abstract}
\textbf{
Distributed quantum computing (DQC) is being actively investigated as a means of scaling the number of qubits across multiple connected quantum devices. This includes quantum circuit compilation and execution management on multiple quantum devices in the network. The latter aspect is very challenging because, while reducing the makespan of job batches remains a relevant objective, novel quantum-specific constraints must be considered, including QPU utilization, non-local gate rate, and the latency associated with queued DQC jobs. In this work, a range of scheduling strategies is proposed, simulated, and evaluated, including heuristics that prioritize resource maximization for QPU utilization, node selection based on heterogeneous network connectivity, asynchronous node release upon job completion, and a scheduling strategy based on reinforcement learning with proximal policy optimization. These approaches are benchmarked against traditional FIFO and LIST schedulers under varying DQC job types and network conditions for the allocation of DQC jobs to devices within a network.
}
\end{abstract}


\maketitle

\section{Introduction}
\label{intro}
Distributed quantum computing (DQC) scales the qubit capacity of quantum systems by interconnecting multiple QPUs and enabling collaborative computation. 
Communication between remote quantum processors requires the distribution and consumption of EPR pairs over quantum links before any inter-processor operations can be executed. EPR pairs, also denoted as Bell states, are the maximally entangled quantum states of a two-qubit system (i.e., a quantum mechanical system composed of two interacting two-level subsystems)~\cite{Brunner2014,Amoretti2020}.

Transitioning from monolithic to distributed quantum computing introduces challenges in algorithm partitioning and execution management~\cite{caleffi2024distributed}. A quantum compiler is responsible for partitioning a monolithic algorithm into local and remote operations, as in~\cite{ferrari2023modular, xu2025remote, cuomo2023optimized}. Execution management, in contrast, focuses on scheduling distributed quantum circuit instances, or jobs,  onto the quantum network for each execution round~\cite{ferrari2024execution, diadamo2021distributed, parekh2021quantum}.

The practical computational advantage of DQC depends on communication overhead, entanglement-generation rate, fidelity, error correction, connectivity, compiler efficiency, algorithmic structure, and execution management performance.

This paper focuses on the execution management of DQC jobs. In classical high performance computing, there is a rich literature of algorithms that aim to reduce the makespan of job batches while maximizing the utilization of computational resources~\cite{johannes2006scheduling,dutton2008parallel,sgall2014multiprocessor}. In the DQC domain, execution management is much more challenging because, while reducing the makespan is still a relevant objective, novel quantum-specific constraints must be taken into account. These constraints are mostly related to the generation and distribution of EPR pairs: there is a tradeoff between rate and fidelity (which measures the quality of a quantum state). Furthermore, there is a time limit for consuming an entangled state before decoherence (i.e., loss of quantum information) due to interactions with the environment.

The main contributions of this paper are as follows:
\begin{itemize}
    \item an integrated DQC simulation framework that models the full DQC workflow, including quantum circuit compilation, job allocation, and execution over heterogeneous quantum processing unit (QPU) networks with varying circuit types and workloads.
    \item the definition and implementation of multiple scheduling strategies for assigning compiled sub-circuits (DQC jobs) to QPU networks for execution. These include three major categories specific to the DQC job scheduling problems: heuristic schedulers (resource-prioritization, EPR-based, and ASAP strategies), a mixed-integer optimization-based scheduler, and a reinforcement learning-based Proximal Policy Optimization (PPO) scheduler.
    \item a comprehensive performance evaluation using multiple metrics, including makespan, QPU utilization, non-local gate rate, system execution latency, and fairness, to assess the performance of the proposed schedulers against the baseline LIST and FIFO schedulers from parallel job scheduling literature.
\end{itemize}

The paper is organized as follows. In Section~\ref{sec:related}, related research on scheduling distributed computation tasks in quantum networks and classical distributed systems is reviewed. Section~\ref{sec:framework} presents the DQC workflow, including the quantum circuit compilation into multiple DQC jobs, the scheduling process, scheduler performance evaluation, and the network model comprising the description of the network topology and time-slot synchronization simulation.
In Section~\ref{sec:performance_metrics}, the performance metrics used to evaluate scheduling methods are introduced. In Section~\ref{sec:algorithms}, multiple scheduling methods for distributing DQC jobs are explained in detail. In Section~\ref{sec:simulation}, the simulation settings and the analysis of the results are presented. Finally, Section~\ref{sec:conlusion} provides the conclusions of the paper.


\section{Related Work}
\label{sec:related}

In classical high performance computing systems, a central challenge is to allocate the resources of parallel machines among competing jobs while satisfying their service requirements \cite{feitelson1995parallel}. System performance is typically evaluated using metrics such as individual and average job completion times, as well as fairness across tasks~\cite{isard2009quincy}.

Job scheduling in classical high performance computing is NP-hard, even when optimizing the preemptive makespan~\cite{johannes2006scheduling}. Common approaches for finding sub-optimal solutions include the use of heuristics~\cite{lifka1995anl,topcuoglu2002performance,turkakin2021comparison} and mixed-integer programming~\cite{tesch2020polyhedral}. However, these techniques cannot be directly applied to distributed quantum computing (DQC), where entanglement resources (e.g., Bell pairs) are limited and susceptible to decoherence. Furthermore, inter-QPU communication, primarily driven by entanglement generation and distribution, contributes substantially to the overall execution time.

Therefore, distributed quantum computing requires dedicated scheduling methods to coordinate the execution of sub-circuits decomposed from a monolithic quantum circuit while accounting for both quantum operation dependencies and network operations required for entanglement generation. Designing effective resource management and scheduling strategies for DQC workloads remains an active research area, with relatively limited work available to date.

Recent work on DQC has increasingly focused on scheduling and resource management. In \cite{xu2025remote}, the authors investigate the Remote Gate Scheduling problem, motivated by the high cost of entangled qubit pairs required for remote quantum gate operations. They consider remote-gate execution paradigms, and propose a hybrid heuristic that dynamically schedules quantum gate operations across distributed QPUs while minimizing entanglement consumption.

Similarly, \cite{chandra2024network} presents a complete DQC workflow, including quantum circuit partitioning, identification of non-local gates, entanglement generation, and the scheduling of both quantum and network operations. The authors compare a resource-constrained project scheduling (RCPSP) formulation with batching against a greedy heuristic. At each scheduling step, the RCPSP approach determines the set of eligible operations based on the availability of quantum resources, such as communication and memory qubits, to improve overall execution efficiency. However, the evaluation is limited to the Quantum Fourier Transform (QFT) algorithm deployed on at most four QPUs.

In \cite{diadamo2021distributed}, the authors study the distributed execution of an accelerated variational quantum eigensolver (VQE) across multiple QPUs and deeper quantum circuits. Their scheduling problem focuses on allocating Ansatz states among distributed QPUs. The authors  consider both a greedy strategy that fills each QPU with as many Ansatz states as possible and a constraint-programming approach. 

Considering the diversity in the types and depths of arriving quantum circuits, effective parallelization of DQC jobs is essential. In~\cite{parekh2021quantum}, the authors develop a mathematical framework for parallel and distributed quantum algorithms and propose a scheduling approach for programs executed across networks of distributed quantum processors. Similarly,~\cite{orenstein2024qgroup} introduces a scheduling algorithm that parallelizes a queue of quantum circuits with heterogeneous depths to minimize the makespan
and maximize the fidelity for each job. The authors benchmark their method against three baseline algorithms and align the number of shots per job to reduce overhead. Their simulation results are based on calibration data from existing IBM quantum computers with varying fidelities.

In this paper, we model the complete DQC workflow, allowing performance evaluation based on the full execution of quantum and network operations rather than on statistical datasets. The network model captures multiple time slots, each associated with heterogeneous queues of arriving quantum circuits that differ in type and depth. We further develop and implement several scheduling strategies specific to DQC job assignment and benchmark them against the FIFO and list-scheduling heuristics used in~\cite{ferrari2024execution}.


\section{DQC Framework}
\label{sec:framework}
In this section, the DQC workflow and network model, including network link settings and synchronized network simulation, are presented.

\subsection{DQC Workflow}
\label{subsec:dqc_workflow}

\begin{figure}
\centering
\includegraphics[width=1\linewidth]{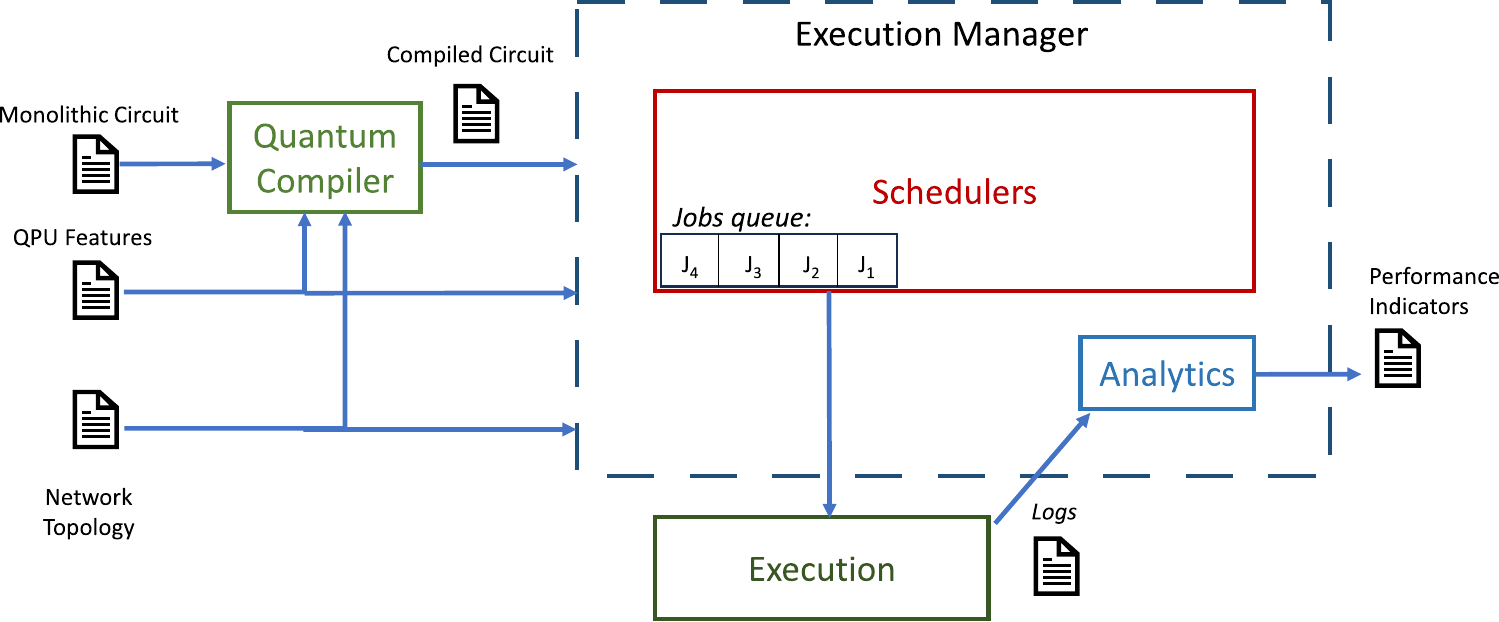} 
\caption{\label{fig:framework} The DQC workflow decomposes a monolithic quantum circuit into sub-circuits (jobs). The execution manager synchronizes multiple schedulers, each independently allocating jobs across the network. The jobs are then executed, and the resulting simulation data are analyzed to assess the performance of each scheduler.}
\hrulefill
\end{figure}

The DQC workflow encompasses circuit compilation, the decomposition of a monolithic quantum circuit into sub-circuits (each treated as an independent job), and the allocation of Quantum Processing Unit (QPU) nodes to execute these jobs, as illustrated in Figure~\ref{fig:framework}.

In Figure~\ref{fig:framework}, the Quantum Compiler, as mentioned in~\cite{ferrari2023modular,bandini2024optimized}, is applied to partition the circuit into executable sub-circuits based on the network topology and QPU capabilities, ensuring that inter-node dependencies are minimized. In contrast, the Scheduler manages the execution of these jobs by organizing the job queue and determining whether they should be executed concurrently or sequentially to achieve optimal performance and resource utilization.

\begin{figure}
\centering
\includegraphics[width=1\linewidth]{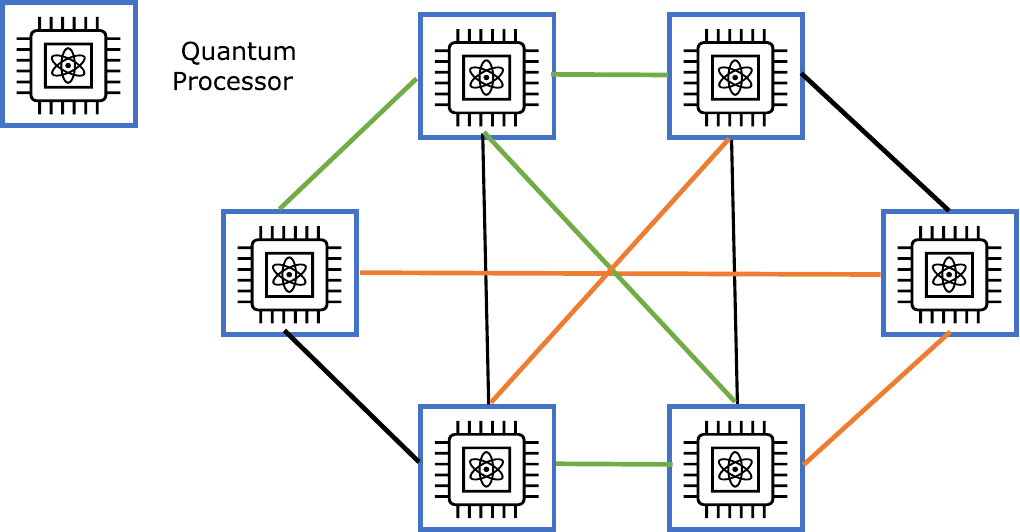}
\caption{\label{fig:qpu_network} Fully connected QPU network with heterogeneous links.}
\hrulefill
\end{figure}

In this work, we consider a fully connected QPU network topology, as it reduces the number of required non-local operations. Each QPU is equipped with data qubits and  communication qubits and is treated as a node in the network. Links between nodes are heterogeneous, as indicated by the different colors in Figure~\ref{fig:qpu_network}.

\subsection{Network model}

In this work, entanglement is assumed to be a resource provided by the network infrastructure as a service to all the nodes. Heterogeneous end-to-end entangled links are modeled using realistic parameters. Including the actual network topology -- which may be possible in a datacenter scenario -- would make the DQC Execution Manager more accurate, but also poorly scalable. Synchronized arrivals of DQC jobs per time slot are assumed.

\subsubsection{Heterogeneous Links}
\label{sub:heterogenous_link}
Links between nodes are characterized by several key parameters, including the entanglement generation cycle time, expected state delay, entanglement success probability, and fidelity. Each of them has an impact on the performance of DQC job execution.

In~\cite{oslovich2025compilation}, the entanglement success probability $P_s$ represents the probability that a single entanglement generation attempt succeeds. For a link of length $d$, it can be modeled as
\begin{equation}
\label{eqn:P_s}
    P_s(d) = \frac{1}{2}\,\eta_{\text{penalty}}
            \left(\eta_{\text{ion}}
            \eta_{FC}^{\text{ion}\rightarrow\text{telecom}}
            \eta_{\text{det}}^{\text{telecom}}
            \right)^2
            10^{-(\alpha/10)(d/2)},
\end{equation}
where $\eta_{\text{ion}}$ is the efficiency for emitting and collecting a photon, $\eta_{FC}^{\text{ion}\rightarrow\text{telecom}}$ is the efficiency of frequency conversion from ion to photons, $\eta_{\text{det}}^{\text{telecom}}$ is the detector efficiency for photons at telecom frequency, $\alpha$ is the attenuation factor, and $\eta_{\text{penalty}}$ accounts for a penalty to truncate the detection window.

The cycle time $t_{\text{cycle}}$ represents the average duration required for a single entanglement generation attempt. Shorter cycle times enable higher repetition rates and, thus, faster entanglement generation.

The state delay $t_{\text{state}}$ represents the expected time to successfully generate an entangled state in a deterministic model and is given by
\begin{equation}
\label{eqn:state_delay}
    t_{\text{state}} = \frac{t_{\text{cycle}}}{P_s}.
\end{equation}
Eq. (\ref{eqn:state_delay}) combines both the time for the entanglement generation attempt and the success probability to quantify the time needed to obtain a usable entangled state.

\subsubsection{Synchronized Network Simulation}
The simulation consists of multiple time slots. In each slot, jobs arrive randomly but remain consistent across different schedulers. The number of arriving jobs follows a truncated Poisson distribution with specified lower and upper bounds. Schedulers immediately receive jobs upon completing the previous queue, allowing the simulation to track execution times and latencies for individual jobs. This setup captures both the stochastic nature of job arrivals and the dynamic resource allocation across the network, enabling a comprehensive evaluation of DQC job scheduling and execution.

\section{Problem Formulation}
\label{sec:problem_formulation}

\subsection{Layers correlations}
\begin{figure}
\centering
\includegraphics[width=1\linewidth]{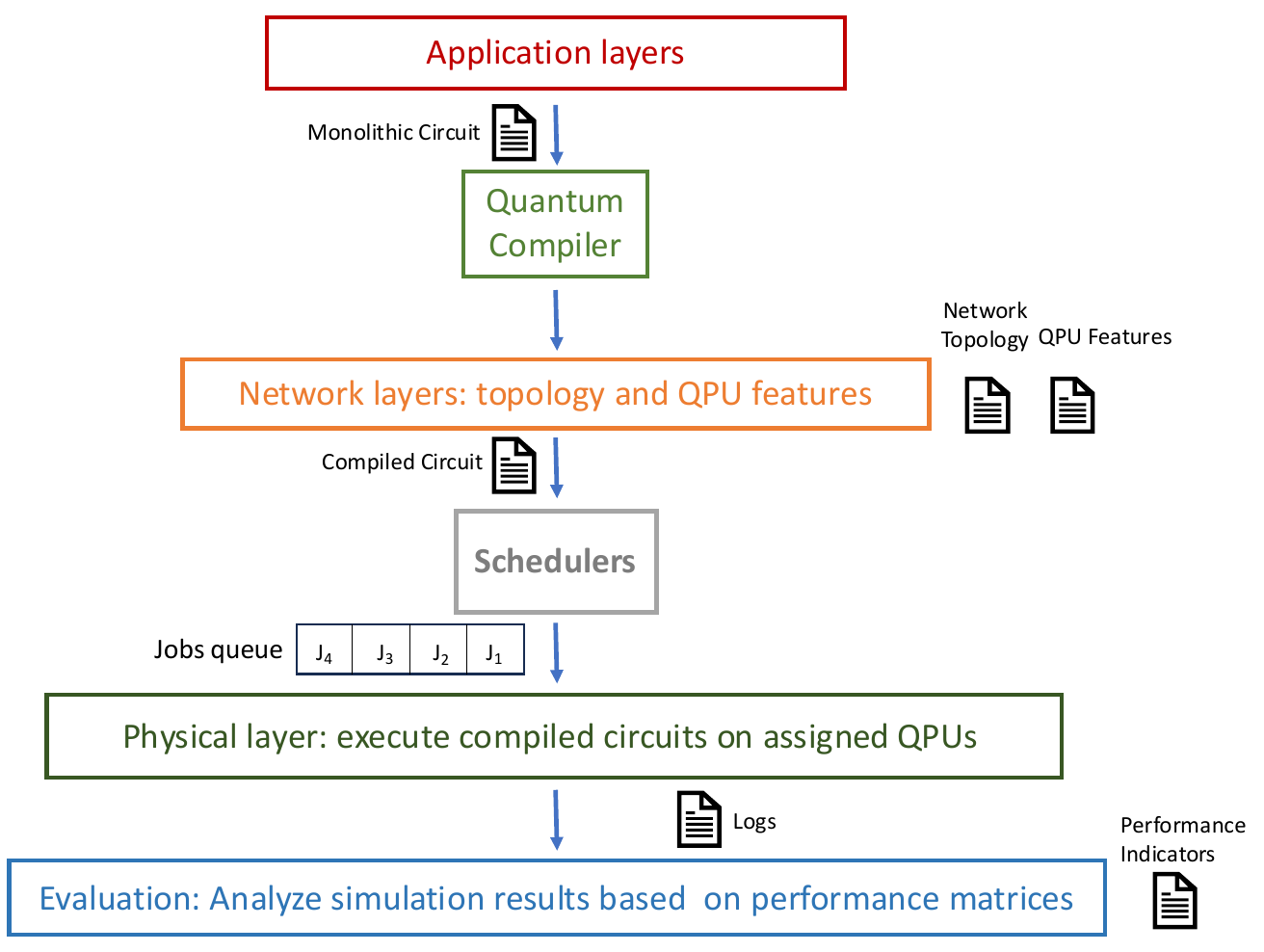}
\caption{\label{fig:layers} DQC Framework layers.}
\hrulefill
\end{figure}

Given that multiple quantum circuits are requested for execution on a fully connected QPU network at the application layer, as illustrated in Fig.~\ref{fig:layers}, the quantum compiler, described in Section~\ref{subsec:dqc_workflow}, collects information about the network topology and QPU capabilities from the network layer and partitions each monolithic circuit into a set of compiled DQC jobs. These jobs are characterized by their required non-local gates, the QPU resources demand, and their estimated execution times, which are calculated by considering both quantum operations and network communication overheads. The detailed execution time estimation follows the same procedure and parameters as in \cite{ferrari2024execution}.

This information is then provided to the scheduler, which determines how the compiled jobs should be assigned to the available QPUs. Once the scheduling decisions, including the placement and start times of the jobs, have been determined, the physical layer performs the simulation by accounting for operation dependencies, execution times, and the associations between QPUs and their connecting links.

After the simulation is completed, the performance of the scheduling algorithms is evaluated using the performance metrics described in Section~\ref{sec:performance_metrics}.

\subsection{Scheduling problem}
\label{subsec:scheduling_problem}

In the DQC scheduling problem, a scheduler assigns quantum processing units (QPUs) to a set of queued jobs $\mathcal{J}$ and determines their execution order over time. Each job $j \in \mathcal{J}$ is characterized by its required non-local gate $E_j$, QPU resource demand $R_j$, and estimated execution time $T_j$.

Two scheduling paradigms are considered: static scheduling and dynamic scheduling.

\paragraph{Static scheduling.}
In the static setting, the QPU network is represented as an graph
\begin{equation}
\mathcal{G} = (\mathcal{N}, \mathcal{E}),
\end{equation}
where $\mathcal{N}$ is the set of QPUs and $\mathcal{E}$ denotes available communication links.

At each scheduling step $t$, the scheduler constructs a mapping
\begin{equation}
\pi_t: \mathcal{A}_t \rightarrow 2^{\mathcal{N}},
\end{equation}
where $\mathcal{A}_t \subseteq \mathcal{J}$ is the set of scheduled jobs at step $t$, and $\pi_t(j)$ denotes the subset of QPUs assigned to job $j$, with $|\pi_t(j)| = R_j$. This assignment must satisfy the resource constraint
\begin{equation}
\bigcup_{j \in \mathcal{A}_t} \pi_t(j) \subseteq \mathcal{N}.
\end{equation}

The resulting static schedule is defined as
\begin{equation}
\mathcal{S}_t = \left\{ \left(j, \pi_t(j), m_j, c_j \right) \mid j \in \mathcal{A}_t \right\},
\end{equation}
where $m_j$ and $c_j$ denote the start and completion times of job $j$ within step $t$.

\paragraph{Dynamic scheduling.}
In the dynamic setting, scheduling evolves over continuous time $t \ge 0$. Each job $j$ is associated with a start time $m_j$ and expected completion time
\begin{equation}
c_j = m_j + T_j.
\end{equation}

At any time $t$, the set of occupied QPUs is
\begin{equation}
\mathcal{B}(t) = \bigcup_{j \in \mathcal{J} : m_j \le t < c_j} \pi_t(j),
\end{equation}
and the available QPUs are given by
\begin{equation}
\mathcal{R}(t) = \mathcal{N} \setminus \mathcal{B}(t).
\end{equation}

Whenever a resource release event occurs (i.e., a job expected completion time $c_j$), the scheduler re-evaluates the remaining unscheduled job set $\mathcal{J}_{\mathrm{rem}}(t)$ and assigns any job $j$ satisfying
\begin{equation}
\pi_t(j) \subseteq \mathcal{R}(t),
\end{equation}

The dynamic schedule is defined as
\begin{equation}
\mathcal{S}_t =
\{ (j, \pi_t(j), m_j, c_j) \mid j \in \mathcal{A}_t \},
\end{equation}
where $\mathcal{A}_t \subseteq \mathcal{J}_{\mathrm{rem}}(t)$ is the set of jobs scheduled at time $t$.


\section{Performance Metrics}
\label{sec:performance_metrics}
The performance metrics described in this section are used to evaluate the DQC job schedulers. Makespan indicates the total time consumed to execute an entire queue of DQC jobs, reflecting how efficiently all jobs are completed. QPU utilization and non-local gate rate measure how effectively the system uses local quantum resources (data qubits) and quantum network resources (EPR pairs). For overall system assessment, execution-latency performance and fairness metrics are also considered to evaluate both latency and fairness among jobs within the queue.

\subsection{Makespan}
\label{sub:makespan}
The makespan $M$ represents the total completion time of a queue of DQC jobs, defined as
\begin{equation}
M = \max_j \left( t_j^{\mathrm{f}} \right) - \min_j \left( t_j^{\mathrm{s}} \right),
\end{equation}
where $t_j^{\mathrm{s}}$ and $t_j^{\mathrm{f}}$ denote the start and finish times of the \textit{j}-th job, respectively.
A shorter makespan indicates that the scheduler completes the DQC jobs more efficiently.

\subsection{QPU Utilization}
The QPU utilization~\cite{ferrari2024execution} measures the proportion of QPU resources occupied by DQC jobs, normalized by the product of the makespan and the total number of available QPUs.
\begin{equation}
U_{\mathrm{QPU}} = \frac{\sum_j p_j q_j}{M n_{\mathrm{QPU}}} \in [0,1],
\end{equation}
where  $p_j$ is the execution time of job $j$, $q_j$ is the number of QPUs required by that job, $n_{\mathrm{QPU}}$ is the total number of QPUs in the system, and $M$ is the makespan. A higher QPUs utilization indicates that the scheduler uses most of the QPUs in the network, which is a desirable feature.

\subsection{Non-Local Gate Rate}

Non-local gate rate quantifies the non-local gate utilization rate for an incoming DQC job queue. For each job, the execution time includes both local and non-local quantum gates. 

It is worth noting that the real bottleneck in DQC is usually link-level EPR generation, EPR-pair consumption, and retry probability. With the exception of EPR-pair consumption, which is a static feature of the job, the others are network features that could be available, but that is not always the case. Conversely, the execution time of the jobs can be measured by the Execution Manager. 
Since non-local gates dominate the execution time (for the reasons listed above), the overlap of non-local gate execution can be approximated using the job execution time.

Let $\mathcal{J}$ denote the arrival set of jobs, where each job $j \in \mathcal{J}$ has a start time $t_j^{\mathrm{s}}$ and a finish time $t_j^{\mathrm{f}}$. Indices $j$ and $k$ represent two distinct jobs in $\mathcal{J}$.

The total pairwise overlapping time is defined as
\begin{equation}
T_{\mathrm{overlap}} =
\sum_{\substack{j,k \in \mathcal{J} \\ j < k}}
\max\!\left(0,\,
\min(t_j^{\mathrm{f}}, t_k^{\mathrm{f}}) -
\max(t_j^{\mathrm{s}}, t_k^{\mathrm{s}})
\right),
\end{equation}
where $j < k$ ensures that each job pair is counted only once.

To normalize the pairwise overlapping time, we define the maximum possible overlap between every pair of jobs as
\begin{equation}
T_{\mathrm{max}} =
\sum_{\substack{j,k \in \mathcal{J} \\ j < k}}
\min\!\left(
t_j^{\mathrm{f}} - t_j^{\mathrm{s}},
t_k^{\mathrm{f}} - t_k^{\mathrm{s}}
\right).
\label{eq:non_local_gate_normalization}
\end{equation}

The bounded network utilization is then
\begin{equation}
U_{g} = \frac{T_{\mathrm{overlap}}}{T_{\mathrm{max}}} \in [0,1].
\end{equation}

A higher $U_{g}$ indicates that more non-local gates are executed simultaneously, reflecting heavier usage of the network's entanglement generation capabilities. However, low values for this metric may reflect poorer parallelism.

\subsection{System Execution-Latency Performance}
The Execution-Latency Performance (ELP) of each job quantifies how close its latency is to the ideal execution time:
\begin{equation}
\text{ELP}_{j} = \frac{e_j}{l_j},
\end{equation}
where $e_j$ is the execution time of job $j$, and the latency $l_j$ is the sum of the execution time and the waiting time for the job in the queue.  
The ideal value of $\text{ELP}_{j} = 1$ indicates that the job is executed without waiting.

The System Execution-Latency Performance (SELP) represents the geometric mean of the proportion of execution time and latency for all the jobs:
\begin{equation}
\text{SELP} = \left( \prod_{j=1}^{|\mathcal{J}|} \text{ELP}_{j} \right)^{\frac{1}{|\mathcal{J}|}},
\end{equation}
where $|\mathcal{J}|$ is the total number of jobs.  
An ideal $\text{SELP} = 1$ signifies that, on average, all jobs achieve their ideal execution times.

\subsection{Fairness Among Jobs}
The fairness $f$ metric measures the execution times and latencies among jobs, capturing how equally the schedulers affect them:
\begin{equation}
f = 1-\sigma(\text{ELP}),
\end{equation}
where $\sigma(\text{ELP})$ denotes the standard deviation of $\text{ELP}_{j}$ for each DQC job $j$.  An ideal value of $f = 1$ indicates perfect fairness—i.e., all jobs are executed concurrently in the quantum network.

\section{Scheduling Methods}
\label{sec:algorithms}
As multiple DQC jobs with varying execution times arrive in each time slot, schedulers are designed to determine their concurrency and execution order. Given that scheduling in a parallelized queue is NP-hard, schedulers aim to deliver solutions based on different optimization objectives. In this work, three scheduler classes are proposed for DQC job coordination: heuristic methods (resource-prioritization, EPR-based, and ASAP strategies), mixed-integer optimization, and reinforcement learning based on Proximal Policy Optimization (PPO).

\subsection{Resource-Priority Scheduler}
As QPU resources are the primary constraint in DQC job scheduling (Section~\ref{subsec:scheduling_problem}), the Resource-Priority Scheduler aims to maximize the utilization of the available QPU resources. Its scheduling decisions are guided by two criteria. First, it prioritizes the subsets of the parallelized jobs that maximize node utilization, ensuring that as many nodes as possible are actively engaged at each scheduling step. Second, among subsets achieving the same maximal utilization, it prioritizes the one with the shortest total estimated execution time. The pseudo-code is described in Algorithm~\ref{alg:resource_prioritize}.

This procedure ensures that the scheduler simultaneously maximizes QPU utilization and minimizes total execution time at each scheduling step until all DQC jobs in the time slot have been scheduled.

\begin{algorithm}
\caption{Resource-Priority Scheduler}
\label{alg:resource_prioritize}
\begin{algorithmic}[1]

\STATE \textbf{Input:} arrival job set $\mathcal{J}$, nodes $\mathcal{N}$
\STATE \textbf{Output:} schedule $\mathcal{S}$

\FORALL{job $j \in \mathcal{J}$}
    \STATE Estimate execution time $T_j$ and required QPUs $R_j$
\ENDFOR

\WHILE{$\mathcal{J} \neq \emptyset$}
    \STATE $\text{best\_combo}\gets \emptyset$, $\text{best\_util}\gets 0$, $\text{best\_time}\gets 0$

    \FORALL{combinations $C$ of $\mathcal{J}$}
        \STATE $R_{\text{tot}}\gets \sum_{j\in C} R_j$
        \IF{$R_{\text{tot}} \le |\mathcal{N}|$}
            \STATE $T_{\text{tot}} \gets \sum_{j\in C} T_j$
            \IF{$R_{\text{tot}} > \text{best\_util}$ \\
                \textbf{or} ($R_{\text{tot}} = \text{best\_util}$ and $T_{\text{tot}} < \text{best\_time}$)}
                \STATE $\text{best\_combo}\gets C$
                \STATE $\text{best\_util}\gets R_{\text{tot}}$
                \STATE $\text{best\_time}\gets T_{\text{tot}}$
            \ENDIF
        \ENDIF
    \ENDFOR
    \FORALL{job $j \in \text{best\_combo}$}
        \STATE Remove job $j$ from $\mathcal{J}$
    \ENDFOR
\ENDWHILE

\STATE \textbf{return} $\mathcal{S}$

\end{algorithmic}
\end{algorithm}

\subsection{EPR Scheduler}
Similar to the earliest finish time heuristic \cite{topcuoglu2002performance} in classical distributed computing, which prioritizes tasks with minimal communication and computation costs, the proposed EPR scheduler follows a similar intuition. In the DQC setting, the communication cost—dominated by entanglement generation—constitutes the majority of the job execution time. 

Therefore, we design our EPR scheduler to prioritize DQC jobs requiring a smaller number of EPR pairs. The job queue is reordered in ascending order based on EPR usage, ensuring that jobs requiring fewer non-local entanglement resources are executed earlier, subject to the constraint of available nodes at each step. A lower demand for EPR pairs generally leads to shorter execution times, thereby reducing the waiting time of subsequent jobs in the queue. The pseudo-code is described in Algorithm~\ref{alg:epr_scheduler}.

In contrast to a Resource-Priority Scheduler, which aims to maximize QPU utilization at every scheduling step, the EPR Scheduler focuses on executing low-EPR jobs early rather than saturating node usage. Consequently, the EPR scheduler has low QPU utilization because its priority scheme terminates parallelization whenever a job cannot join the previously selected high-priority jobs, even if subsequent jobs could be accommodated.

\begin{algorithm}
\caption{EPR Scheduler}
\label{alg:epr_scheduler}
\begin{algorithmic}[1]

\STATE \textbf{Input:} arrival job set $\mathcal{J}$, nodes $\mathcal{N}$
\STATE \textbf{Output:} schedule $\mathcal{S}$

\FORALL{job $j \in \mathcal{J}$}
    \STATE Required QPUs $R_j$, and EPR usage $E_j$
\ENDFOR
\STATE Sort $\mathcal{J}$ in ascending order of $E_j$

\WHILE{$\mathcal{J} \neq \emptyset$}
    \STATE Used node set $U \gets \emptyset$
    \FORALL{job $j \in \mathcal{J}$}
        \STATE combination $\mathcal{C} \gets$ nodes not in $U$
        \IF{$|\mathcal{C}| < R_j$} \STATE \textbf{continue} \ENDIF

        \STATE Remove job $j$ from $\mathcal{J}$
    \ENDFOR
\ENDWHILE

\STATE \textbf{return} $\mathcal{S}$

\end{algorithmic}
\end{algorithm}

\subsection{EPR Scheduler with Node Selection}

Given that the EPR scheduler has relatively low QPU utilization and considering that the network exhibits heterogeneous link characteristics, the integration of a node-selection algorithm (Algorithm~\ref{alg:node_selection}) with the EPR scheduler enhances resource allocation by prioritizing nodes that are connected via the most efficient links to reduce job's execution time and improve overall scheduling performance.

\begin{algorithm}
\caption{Node-selection algorithm}
\label{alg:node_selection}
\begin{algorithmic}[1]
\STATE \textbf{Input:} Quantum network graph $G$ with link weights, select $\mathcal{K}$ nodes in available nodes $\mathcal{N}$
\STATE \textbf{Output:} Optimal node group $N^\ast$

\STATE $N^\ast \gets \text{None}$
\STATE $\text{min\_weight} \gets +\infty$

\FORALL{node subsets $N \subseteq \mathcal{N}$ such that $|N| = \mathcal{K}$}
    \STATE $\text{weight} \gets 0$
    
    \FORALL{pairs $(u,v)$ where $w(u,v)$ is the link weight}
        \STATE $\text{weight} \gets \text{weight} + w(u,v)$
    \ENDFOR
    
    \IF{$\text{weight} < \text{min\_weight}$}
        \STATE $\text{min\_weight} \gets \text{weight}$
        \STATE $N^\ast \gets N$
    \ENDIF
\ENDFOR

\RETURN $N^\ast$
\end{algorithmic}
\end{algorithm}

\subsection{ASAP Scheduler}
As a dynamic scheduling strategy in this paper, and analogous to classical back-filling strategies \cite{lifka1995anl}, the ASAP scheduling method releases nodes asynchronously as their assigned jobs complete. Rather than waiting for all nodes to become idle at the end of each scheduling step, as the previous schedulers do, this scheduler allocates newly freed nodes to pending jobs whenever possible. This dynamic reassignment improves overall node utilization across the network, as described in Algorithm~\ref{alg:asap_scheduler}.

\begin{algorithm}
\caption{ASAP Scheduler with Node Release}
\label{alg:asap_scheduler}
\begin{algorithmic}[1]

\STATE \textbf{Input:} arrival job set $\mathcal{J}$, nodes $\mathcal{N}$
\STATE \textbf{Output:} schedule $\mathcal{S}$

\STATE For each node $n \in \mathcal{N}$ set $available\_time[n] \gets 0$
\STATE $\mathcal{S} \gets \emptyset$, $t \gets 0$

\FORALL{job $j \in \mathcal{J}$}
    \STATE Estimate execution time $T_j$ and required QPUs $R_j$
\ENDFOR

\WHILE{$\mathcal{J} \neq \emptyset$}

    \STATE $free\_nodes \gets \{n \in \mathcal{N} \mid available\_time[n] \le t\}$

    \FORALL{job $j \in \mathcal{J}$}
        \IF{$|free\_nodes| \ge R_j$}
            \STATE Select $R_j$ nodes from $free\_nodes$
            \STATE $assigned\_nodes \gets$ selected nodes
            
            \FORALL{$n \in assigned\_nodes$}
                \STATE $available\_time[n] \gets t + T_j$
            \ENDFOR
            
            \STATE Remove job $j$ from $\mathcal{J}$
            \STATE Remove $assigned\_nodes$ from $free\_nodes$
        \ENDIF
    \ENDFOR

    \STATE $t \gets \min_{n \in \mathcal{N}} available\_time[n]$ 
    
\ENDWHILE

\STATE \textbf{return} $\mathcal{S}$

\end{algorithmic}
\end{algorithm}

\subsection{MILP Scheduler}

To address the combinatorial complexity of heuristic scheduling, the MILP scheduler formulates the DQC job allocation problem as a time-indexed Resource-Constrained Project Scheduling Problem (RCPSP). In contrast to heuristic Resource-Priority strategies, it computes a globally optimal schedule by minimizing the total makespan under explicit QPU availability constraints.

Each job $j \in \mathcal{J}$ is associated with an estimated execution time $T_j$ and required number of QPUs $R_j$. A discrete time horizon $H$ is defined as:
\begin{equation}
H = \sum_{j \in \mathcal{J}} T_j
\end{equation}

The scheduler introduces binary decision variables $x_{j,t}$, defined as:
\begin{equation}
x_{j,m} =
\begin{cases}
1 & \text{if job } j \text{ starts at time } m_j \\
0 & \text{otherwise}
\end{cases}
\end{equation}

The objective is to minimize the makespan $\min M$. 
subject to the following constraints.

The makespan is defined by the latest finishing job:
\begin{equation}
M \ge (m_j + T_j)\, x_{j,m}, \quad \forall j \in \mathcal{J}, \forall m_j
\end{equation}

Each job must start exactly once:
\begin{equation}
\sum_{t=0}^{H - T_j} x_{j,m} = 1, \quad \forall j \in \mathcal{J}
\end{equation}

The total QPU utilization at any time step $t$ does not exceed the available QPU resources:
\begin{equation}
\sum_{j \in \mathcal{J}} \sum_{\substack{m_j \le t < m_j + T_j}} R_j x_{j,m}
\le |\mathcal{N}|, \quad \forall t \in [0,H]
\end{equation}

After solving the MILP, the start times of all DQC jobs are extracted from the optimal solution, and the MILP scheduler assigns jobs to nodes according to the resulting schedule at each time step.

\begin{algorithm}
\caption{MILP Scheduler}
\label{alg:milp_scheduler_aligned}
\begin{algorithmic}[1]

\STATE \textbf{Input:} job set $\mathcal{J}$, nodes $\mathcal{N}$
\STATE \textbf{Output:} schedule $\mathcal{S}$

\FORALL{job $j \in \mathcal{J}$}
    \STATE Estimate execution time $T_j$ and required QPUs $R_j$
\ENDFOR

\STATE Define horizon $H \gets \sum_{j \in \mathcal{J}} T_j$

\STATE Build MILP with variables $x_{j,m}$ and $M$

\STATE Minimize $M$

\STATE subject to:
\STATE \hspace{0.5cm} each job starts once
\STATE \hspace{0.5cm} makespan consistency constraints
\STATE \hspace{0.5cm} nodes constraints

\STATE Solve MILP

\STATE Initialize schedule $\mathcal{S} \gets \emptyset$

\FORALL{job $j \in \mathcal{J}$}
    \STATE Extract start time $m_j$ from $x_{j,m}$
    \STATE Add $(j, m_j, T_j)$ to $\mathcal{S}$
\ENDFOR

\STATE Sort $\mathcal{S}$ by start time for multiple steps of the Scheduler

\STATE \textbf{return} $\mathcal{S}$

\end{algorithmic}
\end{algorithm}

\subsection{PPO Scheduler}

As an example of a learning-based scheduler for the DQC problem, the Proximal Policy Optimization (PPO) algorithm, an actor--critic method~\cite{schulman2017proximal}, is employed with a specifically designed environment, input space, action space, and reward function to demonstrate the potential of learning-based approaches for scheduling DQC jobs.

PPO is suitable for this setting because it supports discrete action spaces, which naturally align with the job-selection process at each scheduling step. Moreover, PPO is a widely adopted policy-gradient method due to its implementation simplicity, training stability, and relatively low sensitivity to hyperparameter tuning compared with earlier policy optimization approaches such as Trust Region Policy Optimization (TRPO)~\cite{schulman2017proximal}. Its effectiveness with minimal hyperparameter tuning has also been demonstrated across a variety of reinforcement learning benchmarks~\cite{yu2022surprising}.

\subsubsection{Input State}
\label{sec:input_space}

The input state is represented by a matrix $S \in \mathbb{R}^{|\mathcal{J}| \times 4}$, encoding the features of $|\mathcal{J}|$ DQC jobs. Each row $s_j$ corresponds to job $j$ and is characterized by four features: the required number of QPUs ($R_j$), the number of EPR pairs ($E_j$), the estimated execution time ($T_j$), and a binary mask indicator ($B_j \in \{\text{True}, \text{False}\}$). Formally, the state matrix is defined as:

\begin{equation}
S = \begin{bmatrix} 
s_1 \\ 
s_2 \\ 
\vdots \\ 
s_{|\mathcal{J}|} 
\end{bmatrix} = 
\begin{bmatrix} 
R_1 & E_1 & T_1 & B_1\\ 
R_2 & E_2 & T_2 & B_2\\ 
\vdots & \vdots & \vdots & \vdots\\ 
R_{|\mathcal{J}|} & E_{|\mathcal{J}|} & T_{|\mathcal{J}|} & B_{|\mathcal{J}|} 
\end{bmatrix}
\end{equation}

The input state is updated after selecting a job by masking the selected job and any remaining queued jobs that cannot be executed concurrently due to insufficient QPU availability.

\subsubsection{Action}

At each scheduling step, the actor network produces a logit vector
$\boldsymbol{\ell}=[\ell_1,\ldots,\ell_{|\mathcal{J}|}]$ for all jobs.
Invalid jobs are masked according to the job-status vector, yielding
masked logits $\tilde{\ell}_j$. The probability of selecting job $j$
is computed using a temperature-scaled softmax policy.

\begin{equation}
\pi_{\theta}(a_j|S)
=
\frac{\exp(\tilde{\ell}_j/\kappa)}
{\sum_{i=1}^{|\mathcal{J}|}\exp(\tilde{\ell}_i/\kappa)},
\end{equation}

where $\kappa$ controls the exploration level (we set $\kappa = 1$).

To enable stochastic multi-job selection, the Gumbel-Top-$K$ procedure \cite{kool2019stochastic} is applied to produce a ranking in which
jobs with higher policy probabilities are more likely to be selected
while maintaining exploration. 

The selected jobs are executed in parallel and removed from the
queue. The action-selection procedure is repeated until all
jobs have been scheduled.

\subsubsection{Reward}

The reward function encodes the policy to minimize the makespan while preserving potential spare node resources, allowing for further optimization of execution time through the node selection algorithm (Algorithm \ref{alg:node_selection}).

\paragraph{Normalized latency penalty term}

Since the number and types of arriving DQC jobs in each time slot are stochastic, directly using makespan as the reward function is not applicable. Instead, the adopted strategy is to minimize the latency of each job. 

The $l_j$ is the latency of job $j \in \mathcal{J}$. Let $e_{(j)}$ denote execution times sorted in descending order. The normalized latency penalty is defined as
\begin{equation}
R_{\text{lat}} =
\frac{\sum_{j=1}^{|\mathcal{J}|} l_j}
{\sum_{j=1}^{|\mathcal{J}|} (|\mathcal{J}| - j + 1)\, e_{(j)}}.
\end{equation}

This term penalizes schedules that lead to high job latency, where the maximum latency across all jobs defines the makespan. In this way, both latency and the actual execution time observed during training are leveraged to minimize the makespan.

\paragraph{EPR-aware incentive term}

The EPR-based reward is defined as
\begin{equation}
\label{eqn:epr_node_selection_reward}
\begin{aligned}
R_{\mathrm{EPR}}
&=
\frac{1}{|\mathcal{J}|}
\sum_{\upsilon}\sum_{j}
\frac{1}{\beta E_j (o_j + 1) + \gamma D_{\upsilon-1}}, \\
D_{\upsilon-1}
&= \max_{j} E_j,
\qquad
D_0 = 0.
\end{aligned}
\end{equation}

The coefficient $\gamma$ weights the EPR contribution from the previous step, represented by $D_{\upsilon-1}$. Maximizing this reward term encourages the jobs combined in earlier steps to have the smallest possible maximum EPR requirement, thereby minimizing their latency influence on subsequent steps.

Within a parallel group, the coefficient $\beta$ weights the term $E_j(o_j+1)$, where $E_j$ denotes the EPR requirement of job $j$, and $o_j \in \{0,\ldots,\upsilon-1\}$ represents its node-selection priority among the $\upsilon$ parallel jobs. Minimizing $\sum_j E_j(o_j+1)$ promotes assigning jobs with larger EPR requirements to nodes connected by higher-quality links, thereby further reducing the execution time compared to assignments over low-quality links.

\paragraph{Integrated reward}
The integrated reward used for PPO with node selection is defined as

\begin{equation}
\label{eqn:ppo_node_selection_reward}
r = -R_{\text{lat}} + \iota R_{\text{EPR}},
\end{equation}

where $\iota$ balances the normalized latency penalty with an EPR-aware incentive. In this model, we set $\gamma = 0.8$, $\beta = 0.5$, and $\iota = 0.3$.

\subsubsection{Loss Function}

The loss function $\mathcal{L}$ is a weighted sum of the clipped policy loss $\mathcal{L}_{\pi}$, value loss $\mathcal{L}_{v}$, and entropy bonus $\mathcal{H}$, as described in~\cite{schulman2017proximal}:
\begin{equation}
\mathcal{L} = \mathcal{L}_{\pi} + c_v \, \mathcal{L}_{v} - c_e \, \mathcal{H},
\end{equation}
where $c_v$ and $c_e$ are the coefficients of value loss $\mathcal{L}_{v}$ and entropy bonus $\mathcal{H}$, respectively.

\subsubsection{Training}

The PPO model adopts an actor--critic architecture. The actor encodes each job independently using a two-layer multilayer perceptron (MLP) with a hidden dimension of $64$ and ReLU activations, followed by a linear layer that outputs one action logit per job. The critic processes a flattened state representation through a single-hidden-layer MLP with a hidden dimension of $64$ and ReLU activation, followed by a linear value head for state-value estimation. Training uses a mini-batch size of $64$, with policy updates every $1024$ collected transitions.  A discount factor of $0.99$ and a smoothing parameter $0.95$ are used for Generalized Advantage Estimation. The policy uses a clipping ratio of $0.2$, value coefficient $c_v$ of $0.5$ and an entropy regularization coefficient $c_e$ of $0.01$.The learning rate is annealed using a cosine annealing schedule, decreasing from $10^{-3}$ to $10^{-7}$.

Training datasets are generated using the Qoala Simulator \cite{van2025qoala} on the same DQC job types and network topology as the evaluation settings. Separate PPO models are trained for job arrival rates ranging from $4$ to $15$ jobs per time slot, and the scheduler selects the corresponding model according to the observed arrival rate during simulation. Models trained on lower arrival rates typically converge faster; however, all models reach convergence roughly $100$ training epochs.

\subsection{Baseline Schedulers}

Below, we provide a brief explanation of the baseline schedulers from~\cite{ferrari2024execution}.

\subsubsection{FIFO Scheduler}

In the FIFO scheduler, jobs are scheduled strictly in the order of arrival, and each job is assigned to execute when a sufficient number of QPU nodes are available.

\subsubsection{LIST Scheduler}

The LIST scheduler~\cite{johannes2006scheduling} initially follows a FIFO ordering of jobs. However, it enables a job to be scheduled earlier when the required nodes are available, even if its preceding job cannot run in parallel with the already selected jobs, thereby enhancing node utilization and reducing overall waiting time.

\subsection{Schedulers Summary}

As a heterogeneous network is assumed, in this paper, a natural question arises: should links with shorter entangled state preparation and distribution time, be utilized more often, or should all nodes be utilized irrespective of link quality? The proposed Resource-Priority, MILP, and ASAP schedulers aim to exploit all available nodes, leaving no opportunity for node selection. In contrast, the EPR and PPO schedulers allow for explicit node selection. LIST and FIFO serve as our baseline schedulers.

In summary, as shown in Table~\ref{tab:scheduler_summary}, the scheduling strategies differ in queuing flexibility, node utilization, the parallelized job set end condition, and the update mechanism. 

\begin{table*}
\centering
\begin{tabular}{|>{\centering\arraybackslash}m{1.5cm}|>{\centering\arraybackslash}m{3cm}|c|c|>{\centering\arraybackslash}m{3cm}|>{\centering\arraybackslash}m{3cm}|}
\hline
\textbf{Scheduler} & \textbf{Features} & \textbf{Order} & \textbf{Nodes Usage} & \textbf{End Condition} & \textbf{Update} \\ \hline

FIFO & Combine in sequence & Fixed & Low & Next job cannot be grouped & End of parallelized job set \\ \hline

LIST & Allow later jobs to combine & Flexible & Medium & No jobs or no available nodes & End of parallelized job set \\ \hline

Resource-Priority & Prioritize jobs with maximum node usage and minimal estimated time & Flexible & High & No jobs or no available nodes & End of parallelized job set \\ \hline

EPR & Prioritize jobs with minimal EPR & Fixed & Low & Next job cannot be grouped & End of parallelized job set \\ \hline

EPR (Node Selection) & Same as EPR but integrated nodes selection & Fixed & Low & Next job cannot be grouped & End of parallelized job set \\ \hline

MILP & Compute combination to minimize makespan & Flexible & High & No jobs or no available nodes & End of parallelized job set \\ \hline

PPO (Node Selection) & Scheduled by trained models with reward & Flexible & Medium & No jobs or no available nodes & End of parallelized job set \\ \hline

ASAP & Jobs assigned dynamically & Flexible & High & No jobs or no available nodes & When nodes available \\ \hline

\end{tabular}
\caption{Summary table of the schedulers.}
\label{tab:scheduler_summary}
\hrulefill
\end{table*}

\section{Simulation Evaluation}
\label{sec:simulation}

Simulations have been developed and executed using Qoala~\cite{van2025qoala}, which provides an execution environment for quantum network nodes, modeling the node’s classical and quantum components and the runtime coordination between them. It represents programs in a form that combines host code, local quantum routines, and entanglement-request routines, and it uses a task-based scheduler to manage dependencies, asynchronous execution, and the interaction between classical control and quantum operations. The Qoala simulator offers a user-friendly Python API that allows users to mimic both the software and hardware of real quantum network nodes, execute quantum programs, and gather statistics.

\subsection{Simulation Settings}

To evaluate the sensitivity of circuit-set selection, we vary the sampling distribution of circuits. Moreover, we consider various network-load settings to assess scheduler performance under short and long queues of compiled DQC jobs. In these scenarios, a larger number of compiled jobs arises either from decomposing large quantum circuits into multiple sub-jobs during compilation, resulting in longer queues, or from the presence of many small circuits.

\subsubsection{DQC Jobs}

To compare the scheduling algorithms presented in Section~\ref{sec:algorithms}, we consider several quantum circuits with $5$ to $15$ qubits. As large quantum circuits can be decomposed into multiple DQC jobs, we avoid scaling the number of qubits to excessively large values so as not to impose pressure on the circuit compilation process. Instead, we increase the arrival number and diversity of quantum circuit types in the job queue. These circuits are of practical interest and are briefly described in Appendix ~\ref{sec:apendix_DQC_job_types}. The types of these quantum circuits (DQC jobs) include the preparation of GHZ and graph states, as well as the execution of the Quantum Approximate Optimization Algorithm (QAOA), Quantum Fourier Transform (QFT), and Variational Quantum Eigensolver (VQE).

\subsubsection{Job Selection Distribution}

A set of jobs $\Gamma$, comprising all types of quantum circuits and used for random selection, is ordered by the number of EPR pairs required, from smallest to largest. In the simulations, both the network load and the job selection distribution are varied. 

A non-uniform sampling distribution is imposed over the ordered set of jobs. Each job $j$ is assigned a weight
\begin{equation}
\label{eqn:weight}
    w_j = j^{\tau}, \qquad j = 1, \dots, |\Gamma|,
\end{equation}
where $\tau \in [0,1]$ controls the degree of bias. When $\tau = 0$, the distribution is uniform, whereas $\tau = 1$ produces a linear weighting. The sampling probabilities $\xi$ are obtained by normalizing these weights:
\begin{equation}
\label{eqn:sample_pro}
    \xi_j = \frac{w_j}{\sum_{j=1}^{|\Gamma|} w_j}.
\end{equation}

As the bias $\tau$ in the job selection distribution increases, jobs requiring a larger number of EPR pairs become more likely to be sampled.

\subsubsection{Link Parameters}

Following the heterogeneous links modeled in Section~\ref{sub:heterogenous_link}, and according to~\cite{oslovich2025compilation}, the link parameters are summarized in Table~\ref{tab:link_parameter}, where $\eta_{\text{penalty}} = 0.12$ corresponds to a fidelity of $0.88$, and $\eta_{\text{penalty}} = 0.20$ corresponds to a fidelity of $0.95$. We recall that the fidelity $F$ quantifies the quality of the generated entangled state relative to an ideal Bell pair, with higher fidelity indicating a state closer to the desired maximally entangled state.

\begin{table*}
    \centering
    \setlength{\tabcolsep}{5pt}
    \begin{tabular}{l m{1.5cm} m{1.5cm} m{1.5cm} m{0.8cm} m{0.8cm} m{0.6cm}}
        \hline
        Link quality & $\eta_{FC}^{\text{ion}\rightarrow\text{telecom}}$ & $\eta_{\text{det}}^{\text{telecom}}$ & $\eta_{\text{penalty}}$ & $\eta_{\text{ion}}$ & $\alpha$ & $d(km)$\\
        \hline
        Bad    & $0.5$ & $0.75$ & $0.12$ & $0.87$ & $0.2$ & $0.1$\\
        Medium & $0.5$ & $0.75$ & $0.20$ & $0.87$ & $0.2$ & $0.1$\\
        Good   & $0.7$ & $0.90$  & $0.20$ & $0.87$ & $0.2$ & $0.1$\\
        \hline
    \end{tabular}
    \caption{Link parameters for a trapped-ion link at $0.1$ km. Parameters are chosen within reasonable ranges as mentioned in~\cite{oslovich2025compilation}}
    \label{tab:link_parameter}
    \hrulefill
\end{table*}

Table~\ref{tab:trapped_ion_link} summarizes the performance parameters for trapped-ion links, including the calculated success probability $P_s$, the corresponding fidelity $F$ of the generated entangled pairs, cycle time $t_{\text{cycle}}$, and the state delay $t_{\text{state}}$, which is derived from $t_{\text{cycle}}$ and $P_s$ according to Eq.~(\ref{eqn:state_delay}).

\begin{table*}
    \centering
    \footnotesize
    \setlength{\tabcolsep}{3pt}
    \begin{tabular}{m{2cm} m{2cm} m{2cm} m{2cm} m{1.5cm}}
        \hline
        Link quality & $t_{\text{cycle}}$ (ns) & $P_s$ & $F$ & $t_{\text{state}}$ (ns) \\
        \hline
        Worst & $1.8 \times 10^{6}$ & $6.37 \times 10^{-3}$ & 0.88 & $2.83 \times 10^{8}$ \\
        Medium & $1.0 \times 10^{6}$ & $1.06 \times 10^{-2}$ & 0.95 & $9.42 \times 10^{7}$ \\
        Good & $2.0 \times 10^{5}$ & $2.99 \times 10^{-2}$ & 0.95 & $6.67 \times 10^{6}$ \\
        \hline
    \end{tabular}
    \caption{Performance parameters for trapped-ion links. $t_{\text{cycle}}$ for the worst-case link quality is estimated based on the experiments in~\cite{Krutyanskiy2023Entanglement}, whereas the lowest $t_{\text{cycle}}$ for a high-quality link uses the optimistic laboratory setting in~\cite{oslovich2025compilation}.}
    \label{tab:trapped_ion_link}
    \hrulefill
\end{table*}

\subsection{Results}

The simulation results for all schedulers, which process arriving DQC jobs over $1000$ time slots, are evaluated using the performance metrics in Section~\ref{sec:performance_metrics}.

\subsubsection{Makespan}

In Table~\ref{tab:makespan_1} and Table~\ref{tab:makespan_2}, as the network load increases and the selection favors DQC jobs requiring more EPR pairs, the makespan, which is the time required to complete all jobs arriving in the time slot, also increases.

\begin{table*}
\centering
\footnotesize
\setlength{\tabcolsep}{6pt}
\begin{tabular}{l c c c c c c c c}
\hline
\textbf{Network settings}
& \multicolumn{2}{c}{\textbf{FIFO}}
& \multicolumn{2}{c}{\textbf{LIST}}
& \multicolumn{2}{c}{\textbf{Resource-Priority}}
& \multicolumn{2}{c}{\textbf{ASAP}} \\
\cline{2-9}
 & Mean & Std & Mean & Std & Mean & Std & Mean & Std \\
\hline
$\lambda = 5$
& 17.5021 & 12.5834
& 17.5233 & 12.5393
& 16.8197 & 11.7641
& 14.7224 & 9.5952 \\
$\lambda = 5$, bias $\tau = 0.5$
& 21.7040 & 14.5638
& 21.4535 & 14.2003
& 20.6842 & 12.9211
& 18.2697 & 10.9104 \\
$\lambda = 10$
& 31.4975 & 17.5949
& 31.1061 & 17.5493
& 28.6669 & 16.4242
& 25.2182 & 13.2191 \\
$\lambda = 10$, bias $\tau = 0.5$
& 39.7166 & 20.5105
& 39.2626 & 20.3510
& 35.7745 & 18.8619
& 32.7642 & 15.2941 \\
\hline
\end{tabular}
\caption{Mean and standard deviation of Makespan (s) under different network settings.}
\label{tab:makespan_1}
\end{table*}

\begin{table*}
\centering
\footnotesize
\setlength{\tabcolsep}{5pt}
\begin{tabular}{l c c c c c c c c}
\hline
\textbf{Network settings}
& \multicolumn{2}{c}{\textbf{EPR}}
& \multicolumn{2}{c}{\textbf{EPR (Node Selection)}}
& \multicolumn{2}{c}{\textbf{MILP}}
& \multicolumn{2}{c}{\textbf{PPO (Node Selection)}} \\
\cline{2-9}
 & Mean & Std & Mean & Std & Mean & Std & Mean & Std \\
\hline
$\lambda = 5$
& 17.9089 & 12.5614
& 9.9353 & 9.4343
& 14.2915 & 10.0121
& 8.8504 & 8.5414 \\
$\lambda = 5$, bias $\tau = 0.5$
& 21.4648 & 13.3939
& 12.1918 & 10.3600
& 17.6049 & 10.9521
& 11.0687 & 9.6266 \\
$\lambda = 10$
& 30.4938 & 17.1863
& 16.6919 & 12.6646
& 25.2211 & 13.9956
& 15.6643 & 12.0537 \\
$\lambda = 10$, bias $\tau = 0.5$
& 38.4979 & 18.6104
& 21.7032 & 14.2917
& 32.5498 & 16.3043
& 20.5242 & 13.3646 \\
\hline
\end{tabular}
\caption{Mean and standard deviation of Makespan (s) under different network settings.}
\label{tab:makespan_2}
\end{table*}

\begin{figure*}
    \centering
    \begin{subfigure}[b]{0.48\linewidth}
        \centering
        \includegraphics[width=\linewidth]{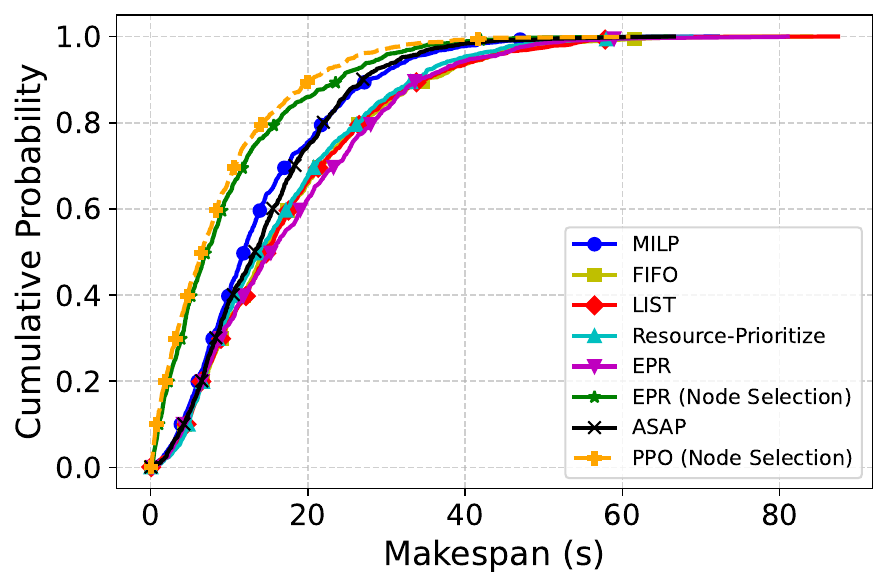}
        \caption{Network load $\lambda = 5$.}
        \label{fig:cdf_makespan}
    \end{subfigure}
    \hfill
    \begin{subfigure}[b]{0.48\linewidth}
        \centering
        \includegraphics[width=\linewidth]{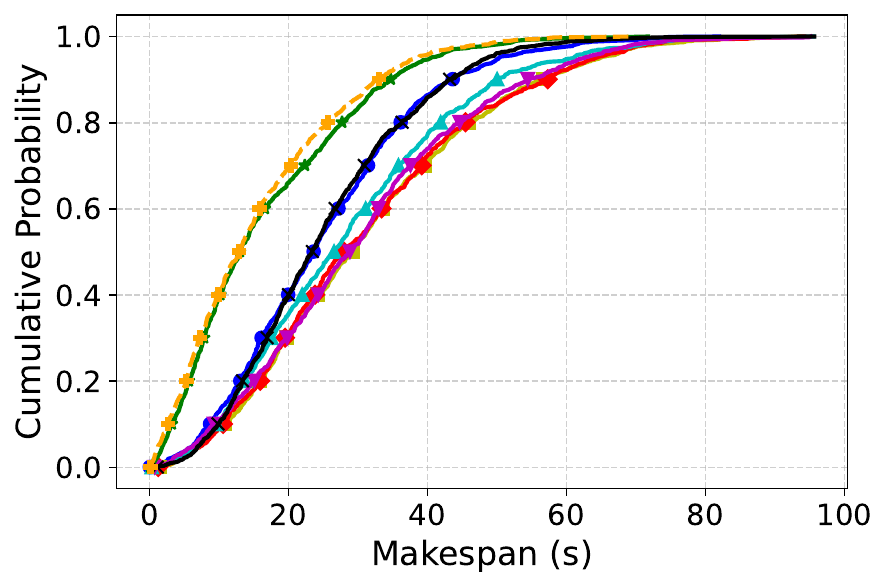}
        \caption{Network load $\lambda = 10$.}
        \label{fig:cdf_8_makespan}
    \end{subfigure}
    
    \begin{subfigure}[b]{0.48\linewidth}
        \centering
        \includegraphics[width=\linewidth]{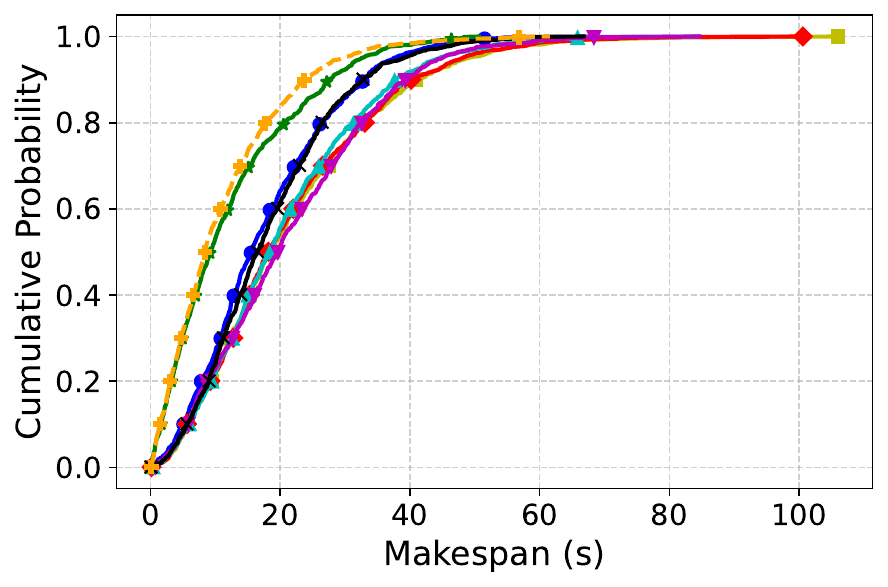}
        \caption{Network load $\lambda = 5$, with biased parameter $\tau = 0.5$ for the job selection.}
        \label{fig:cdf_4_dist_makespan}
    \end{subfigure}
    \hfill
    \begin{subfigure}[b]{0.48\linewidth}
        \centering
        \includegraphics[width=\linewidth]{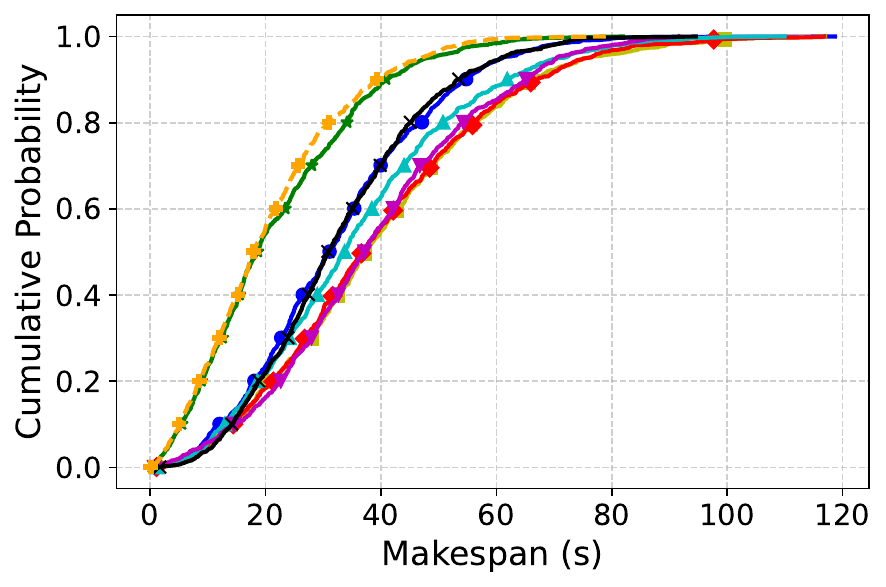}
        \caption{Network load $\lambda = 10$, with biased parameter $\tau = 0.5$ for the job selection.}
        \label{fig:cdf_8_dist_makespan}
    \end{subfigure}
    \caption{The cumulative probability of makespan in each time slot among different schedulers.}
    \hrulefill
    \label{fig:makespan}
\end{figure*}

In Fig.~\ref{fig:makespan}, both EPR (node selection) and PPO (node selection) achieve the lowest makespan across varying network loads and job type distributions compared with other schedulers. This improvement in makespan arises because the node selection algorithm leverages information about heterogeneous network links. Although not all QPUs are fully utilized at every scheduling stage, prioritizing QPUs connected by higher-quality links reduces the time required for entanglement generation and state preparation. In addition, the PPO (node selection) curve remains consistently above that of EPR (node selection), indicating a shorter makespan and highlighting the advantage of learning-based methods in utilizing execution time during training.

The scheduler with the second-lowest makespan is the ASAP and MILP scheduler. The ASAP scheduler allocates QPUs as soon as a DQC job completes and resources become available. This dynamic allocation strategy immediately assigns idle QPUs to waiting jobs, thereby improving resource utilization and reducing the overall makespan. In contrast, the MILP-based scheduler determines a job allocation that minimizes the makespan in advance, based on estimated job execution times.

Among the remaining schedulers, Resource-Priority achieves the third-lowest makespan. EPR, FIFO, and LIST exhibit largely similar makespan performance.

\subsubsection{QPU Utilization}

\begin{figure*}
    \centering
    \begin{subfigure}[b]{0.48\linewidth}
        \centering
        \includegraphics[width=\linewidth]{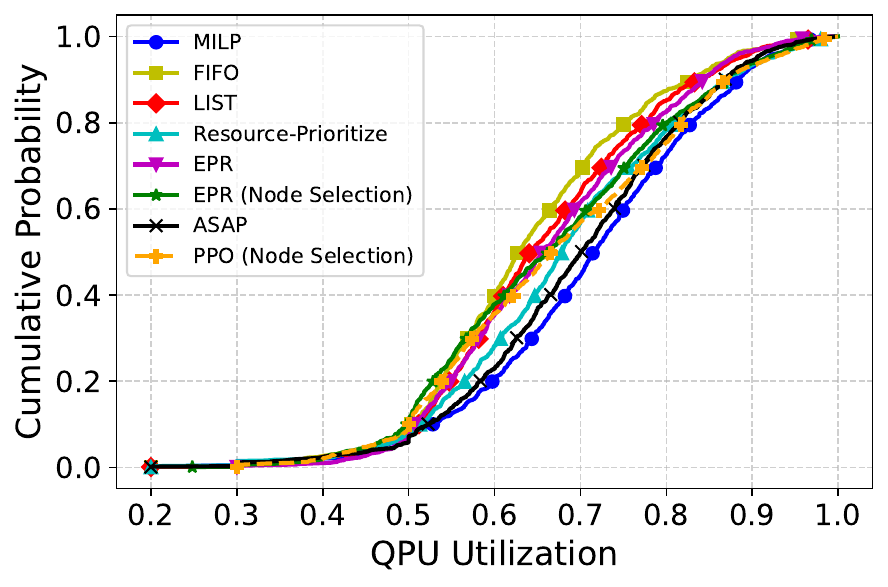}
        \caption{Network load $\lambda = 5$.}
        \label{fig:cdf_qpu_utilization}
    \end{subfigure}
    \hfill
    \begin{subfigure}[b]{0.48\linewidth}
        \centering
        \includegraphics[width=\linewidth]{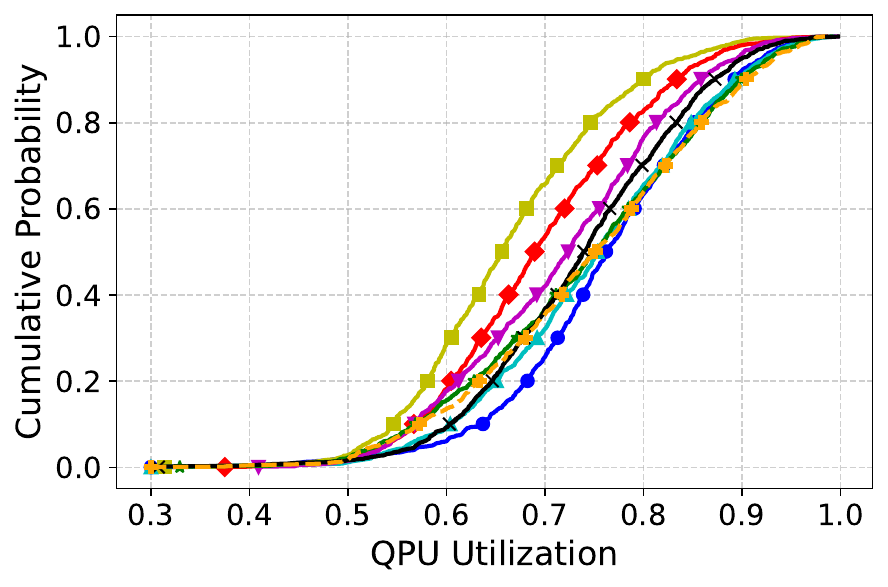}
        \caption{Network load $\lambda = 10$.}
        \label{fig:cdf_8_qpu_utilization}
    \end{subfigure}

    \begin{subfigure}[b]{0.48\linewidth}
        \centering
        \includegraphics[width=\linewidth]{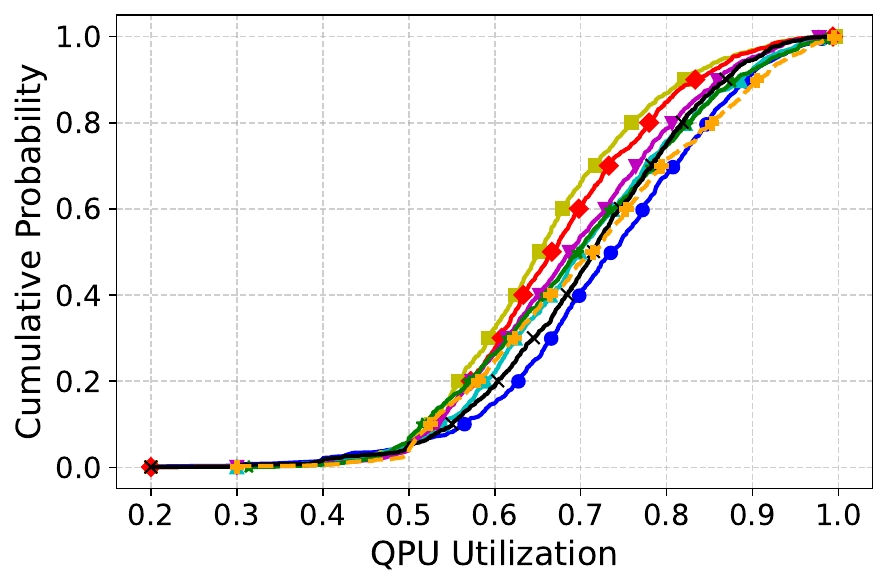}
        \caption{Network load $\lambda = 5$, with biased parameter $\tau = 0.5$ for the job selection.}
        \label{fig:cdf_4_dist_qpu_utilization}
    \end{subfigure}
    \hfill
    \begin{subfigure}[b]{0.48\linewidth}
        \centering
        \includegraphics[width=\linewidth]{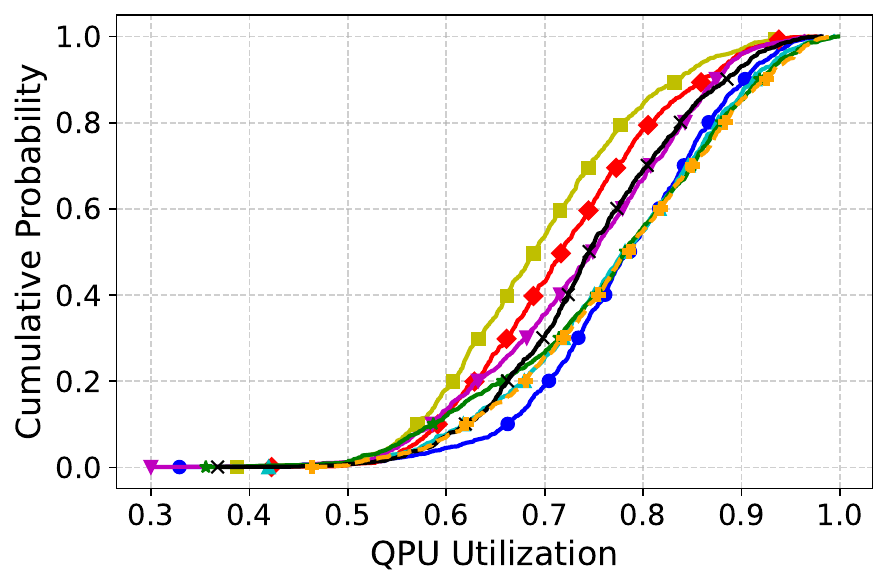}
        \caption{Network load $\lambda = 10$, with biased parameter $\tau = 0.5$ for the job selection.}
        \label{fig:cdf_8_dist_qpu_utilization}
    \end{subfigure}
    \caption{The cumulative probability of QPU utilization in each time slot among different schedulers.}
    \hrulefill
    \label{fig:qpu_utilization}
\end{figure*}

Figure~\ref{fig:qpu_utilization} exhibits the cumulative probability of QPU utilization per time slot under different scheduling methods. As network load increases, the performance differences among the schedulers become more pronounced.

In Table~\ref{tab:qpu_util_1} and Table \ref{tab:qpu_util_2}, FIFO yields the lowest average QPU utilization, whereas MILP, ASAP, and Resource-Priority achieve the highest. The high QPU utilization achieved by these schedulers can be attributed to their strategies for maximizing node utilization.

\begin{table*}
\centering
\footnotesize
\setlength{\tabcolsep}{6pt}
\begin{tabular}{l c c c c c c c c}
\hline
\textbf{Network settings}
& \multicolumn{2}{c}{\textbf{FIFO}}
& \multicolumn{2}{c}{\textbf{LIST}}
& \multicolumn{2}{c}{\textbf{Resource-Priority}}
& \multicolumn{2}{c}{\textbf{ASAP}} \\
\cline{2-9}
 & Mean & Std & Mean & Std & Mean & Std & Mean & Std \\
\hline

$\lambda = 5$
& 0.6438 & 0.1279
& 0.6574 & 0.1302
& 0.6828 & 0.1403
& 0.6961 & 0.1346 \\

$\lambda = 5$, bias $\tau = 0.5$
& 0.6609 & 0.1171
& 0.6740 & 0.1187
& 0.7035 & 0.1310
& 0.7089 & 0.1249 \\

$\lambda = 10$
& 0.6643 & 0.0990
& 0.6956 & 0.1021
& 0.7493 & 0.1119
& 0.7365 & 0.1061 \\

$\lambda = 10$, bias $\tau = 0.5$
& 0.6950 & 0.1002
& 0.7205 & 0.0998
& 0.7738 & 0.1116
& 0.7487 & 0.0993 \\
\hline
\end{tabular}
\caption{Mean and standard deviation of QPU utilization under different network settings.}
\label{tab:qpu_util_1}
\end{table*}

\begin{table*}
\centering
\footnotesize
\setlength{\tabcolsep}{6pt}
\begin{tabular}{l c c c c c c c c}
\hline
\textbf{Network settings}
& \multicolumn{2}{c}{\textbf{EPR}}
& \multicolumn{2}{c}{\textbf{EPR (Node Selection)}}
& \multicolumn{2}{c}{\textbf{MILP}}
& \multicolumn{2}{c}{\textbf{PPO (Node Selection)}} \\
\cline{2-9}
 & Mean & Std & Mean & Std & Mean & Std & Mean & Std \\
\hline

$\lambda = 5$
& 0.6652 & 0.1277
& 0.6670 & 0.1461
& 0.7095 & 0.1371
& 0.6763 & 0.1478 \\

$\lambda = 5$, bias $\tau = 0.5$
& 0.6911 & 0.1256
& 0.6992 & 0.1361
& 0.7291 & 0.1329
& 0.7135 & 0.1404 \\

$\lambda = 10$
& 0.7166 & 0.1095
& 0.7413 & 0.1242
& 0.7606 & 0.1047
& 0.7449 & 0.1234 \\

$\lambda = 10$, bias $\tau = 0.5$
& 0.7382 & 0.1096
& 0.7687 & 0.1227
& 0.7811 & 0.0978
& 0.7780 & 0.1125 \\
\hline
\end{tabular}
\caption{Mean and standard deviation of QPU utilization under different network settings.}
\label{tab:qpu_util_2}
\end{table*}

As the network load increases and DQC jobs become more complex, QPU utilization in MILP and Resource-Priority becomes more evident compared to ASAP. This is because, although ASAP attempts to allocate nodes immediately, it still involves randomness when assigning QPUs to DQC jobs. QPU utilization under both PPO (node selection) and EPR (node selection) increases rapidly as the number of DQC jobs grows.

Among the remaining schedulers, LIST and EPR exhibit largely similar QPU utilization.

\subsubsection{Non-Local Gate Rate}

Non-Local Gate Rate quantifies the overlap in execution times among DQC jobs, which indicates the frequency of usage of non-local gates. A higher overlap in execution time indicates greater reliance on entanglement for the non-local gate. Therefore, lower values in this evaluation are preferable.

However, in the normalization process, the maximum possible overlap between every pair of jobs is computed regardless of whether the jobs can actually be parallelized under the constraint of a fixed number of available nodes per step in Eq.~(\ref{eq:non_local_gate_normalization}). Under higher network load, more incoming jobs cannot be parallelized within the same group. Consequently, jobs from different groups have no real overlap but still contribute to the normalization term.

Similarly, for a given type of quantum circuit, the number of required EPR pairs increases with circuit size, occupying more nodes and forming more parallelized groups, which, in turn, increases the normalization term. As a result, the Non-Local Gate Rate decreases as network load increases and becomes biased toward jobs requiring more EPR pairs, as shown in Table~\ref{tab:net_util_1} and Table \ref{tab:net_util_2}. Therefore, only comparisons between schedulers under the same network conditions (i.e., within each row) remain meaningful.

\begin{table*}
\centering
\footnotesize
\setlength{\tabcolsep}{6pt}
\begin{tabular}{l c c c c c c c c}
\hline
\textbf{Network settings}
& \multicolumn{2}{c}{\textbf{MILP}}
& \multicolumn{2}{c}{\textbf{FIFO}}
& \multicolumn{2}{c}{\textbf{LIST}}
& \multicolumn{2}{c}{\textbf{Resource-Priority}} \\
\cline{2-9}
 & Mean & Std & Mean & Std & Mean & Std & Mean & Std \\
\hline

$\lambda = 5$
& 0.5476 & 0.2532
& 0.4682 & 0.2694
& 0.4843 & 0.2664
& 0.4939 & 0.2456 \\

$\lambda = 5$, bias $\tau = 0.5$
& 0.5298 & 0.2507
& 0.4479 & 0.2707
& 0.4649 & 0.2661
& 0.4957 & 0.2512 \\

$\lambda = 10$
& 0.3220 & 0.1717
& 0.2530 & 0.1641
& 0.2761 & 0.1667
& 0.3195 & 0.1565 \\

$\lambda = 10$, bias $\tau = 0.5$
& 0.2907 & 0.1547
& 0.2435 & 0.1673
& 0.2577 & 0.1661
& 0.2997 & 0.1641 \\
\hline
\end{tabular}
\caption{Mean and standard deviation of non-local gate rate under different network settings.}
\label{tab:net_util_1}
\end{table*}

\begin{table*}
\centering
\footnotesize
\setlength{\tabcolsep}{6pt}
\begin{tabular}{l c c c c c c c c}
\hline
\textbf{Network settings}
& \multicolumn{2}{c}{\textbf{EPR}}
& \multicolumn{2}{c}{\textbf{EPR (Node Selection)}}
& \multicolumn{2}{c}{\textbf{ASAP}}
& \multicolumn{2}{c}{\textbf{PPO (Node Selection)}} \\
\cline{2-9}
 & Mean & Std & Mean & Std & Mean & Std & Mean & Std \\
\hline

$\lambda = 5$
& 0.4692 & 0.2317
& 0.5267 & 0.2602
& 0.5161 & 0.2456
& 0.5268 & 0.2622 \\

$\lambda = 5$, bias $\tau = 0.5$
& 0.4691 & 0.2452
& 0.5385 & 0.2644
& 0.4799 & 0.2405
& 0.5265 & 0.2632 \\

$\lambda = 10$
& 0.3009 & 0.1579
& 0.3626 & 0.1850
& 0.2979 & 0.1525
& 0.3615 & 0.1870 \\

$\lambda = 10$, bias $\tau = 0.5$
& 0.2782 & 0.1591
& 0.3469 & 0.1943
& 0.2614 & 0.1380
& 0.3425 & 0.1937 \\
\hline
\end{tabular}
\caption{Mean and standard deviation of non-local gate rate under different network settings.}
\label{tab:net_util_2}
\end{table*}

\begin{figure*}
    \centering
    \begin{subfigure}[b]{0.48\linewidth}
        \centering
        \includegraphics[width=\linewidth]{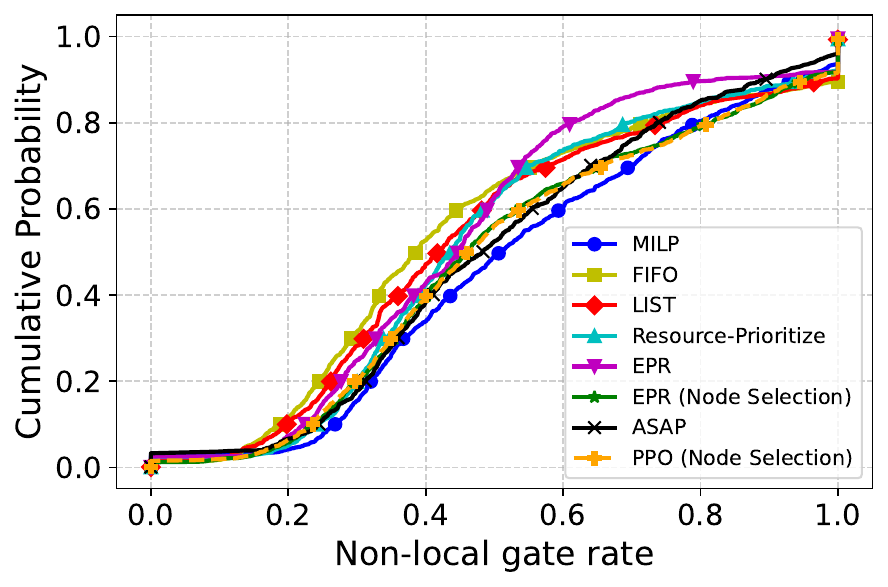}
        \caption{Network load $\lambda = 5$.}
        \label{fig:cdf_non-local_gate_rate}
    \end{subfigure}
    \hfill
    \begin{subfigure}[b]{0.48\linewidth}
        \centering
        \includegraphics[width=\linewidth]{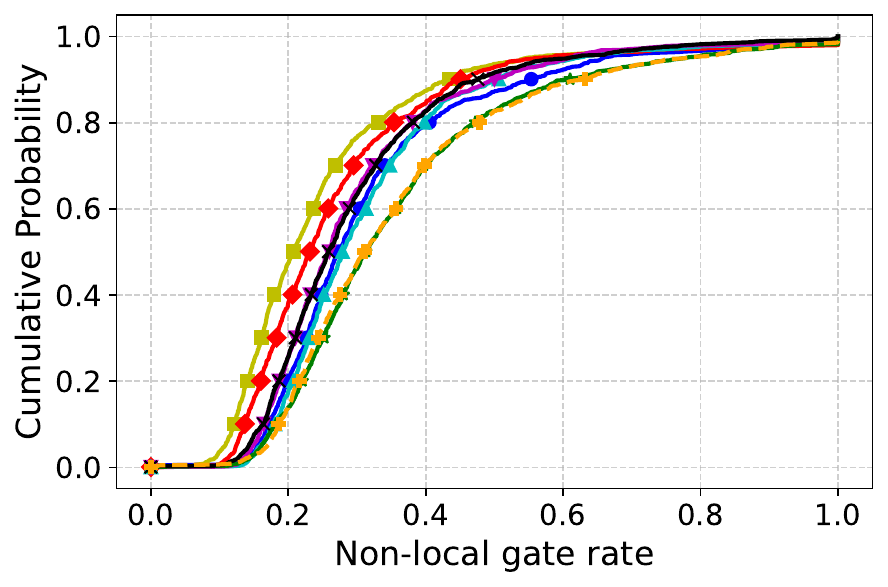}
        \caption{Network load $\lambda = 10$.}
        \label{fig:cdf_8_non-local_gate_rate}
    \end{subfigure}
    
    \begin{subfigure}[b]{0.48\linewidth}
        \centering
        \includegraphics[width=\linewidth]{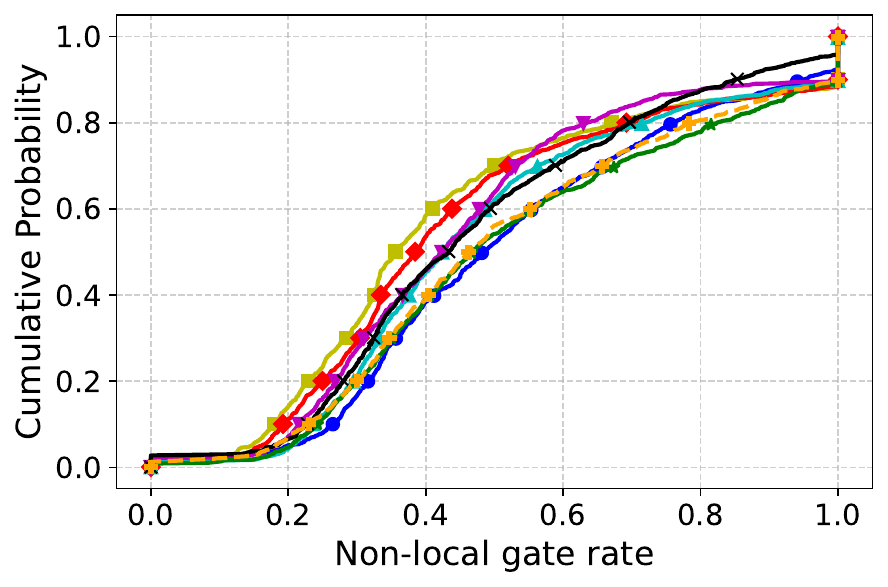}
        \caption{Network load $\lambda = 5$, with biased parameter $\tau = 0.5$ for the job selection.}
        \label{fig:cdf_5_dist_non-local_gate_rate}
    \end{subfigure}
    \hfill
    \begin{subfigure}[b]{0.48\linewidth}
        \centering
        \includegraphics[width=\linewidth]{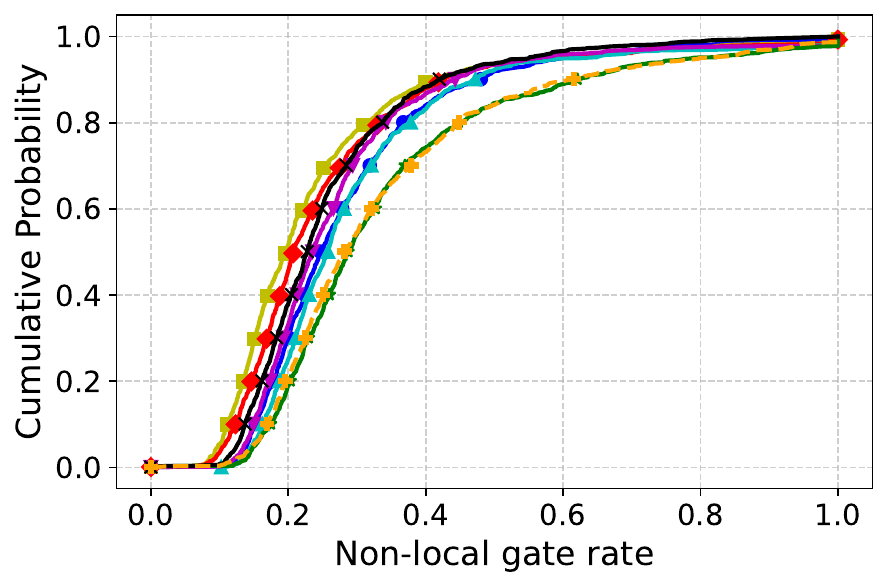}
        \caption{Network load $\lambda = 10$, with biased parameter $\tau = 0.5$ for the job selection.}
        \label{fig:cdf_10_dist_non-local_gate_rate}
    \end{subfigure}
    \caption{The cumulative probability of non-local gate rate in each time slot among different schedulers.}
    \hrulefill
    \label{fig:non_local_gate}
\end{figure*}

When only a few DQC jobs arrive in a time slot, they may all be parallelized, producing complete overlap and a Non-Local Gate Rate of $1$, as shown in Fig.~\ref{fig:non_local_gate}. This effect is particularly evident when the network load is $5$ in Fig.~\ref{fig:cdf_non-local_gate_rate} and Fig.~\ref{fig:cdf_5_dist_non-local_gate_rate}.

Under varying network loads and job distributions, the FIFO scheduler exhibits the lowest Non-Local Gate Rate, as strict request ordering leads to sparse DQC job distributions when grouping. The LIST scheduler consistently produces a slightly higher Non-Local Gate Rate than FIFO.When the network load is low, the EPR scheduler also yields a low Non-Local Gate Rate because it strictly groups jobs by EPR pair requirements. As the number of arriving jobs increases, this constraint becomes less influential since a larger number of jobs increases the likelihood that the next jobs can join previously selected jobs for parallelization.

When the network load is $5$, MILP achieves the highest non-local gate rates, followed by EPR (Node Selection), PPO (Node Selection) and the ASAP scheduler. When the load increases to $10$, the highest values are produced by EPR (Node Selection) and PPO (Node Selection). Under low network load, strategies that utilize more nodes or constantly use the available nodes increase job overlap. However, under higher network load with multiple job groups, EPR (Node Selection) and PPO (Node Selection) tend to schedule jobs with similar execution times within groups, resulting in larger total overlap compared with other schedulers.

For the Resource-Priority scheduler, the number of non-local gates is slightly higher than that of LIST under low network load, but slightly lower than that of the MILP scheduler under high network load.

\subsubsection{System Execution-Latency Performance and Fairness}

Unlike previous evaluations of Makespan, QPU utilization, and Non-Local Gate Rate, which quantify the overall performance of arriving DQC job sets, the System Execution-Latency Performance (SELP) and Fairness metrics assess the performance of individual DQC jobs within the queue arranged by the schedulers.

The Execution-Latency Performance (ELP) measures a job’s actual execution time relative to its latency. Values closer to $1$ indicate shorter waiting times. Fairness is defined with an upper bound of $1$, adjusted by subtracting the standard deviation of the ELP values for all jobs. The optimal value is therefore $1$, which corresponds to minimal variation in ELP among jobs.

However, due to the limited number of QPUs in the network, not all DQC jobs can be executed in parallel. Therefore, both SELP and fairness among jobs are important, with higher values indicating better performance.

\begin{figure*}
    \centering
    \begin{subfigure}[b]{0.48\linewidth}
        \centering
        \includegraphics[width=\linewidth]{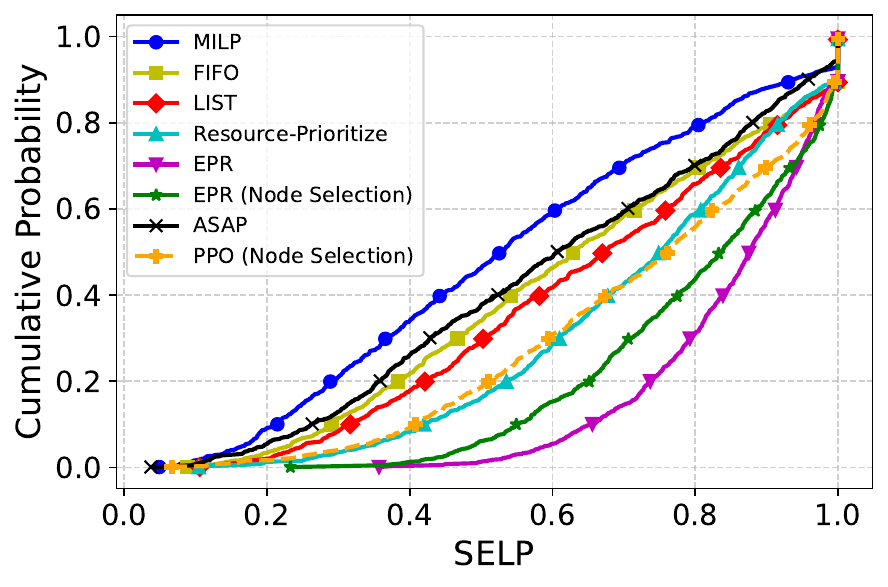}
        \caption{Network load $\lambda = 5$.}
        \label{fig:cdf_selp}
    \end{subfigure}
    \hfill
    \begin{subfigure}[b]{0.48\linewidth}
        \centering
        \includegraphics[width=\linewidth]{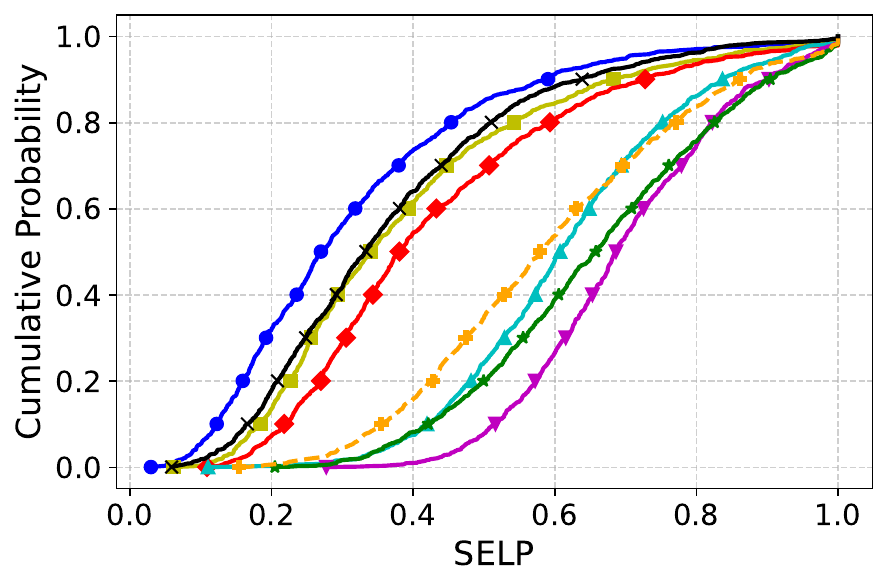}
        \caption{Network load $\lambda = 10$.}
        \label{fig:cdf_10_selp}
    \end{subfigure}
    
    \begin{subfigure}[b]{0.48\linewidth}
        \centering
        \includegraphics[width=\linewidth]{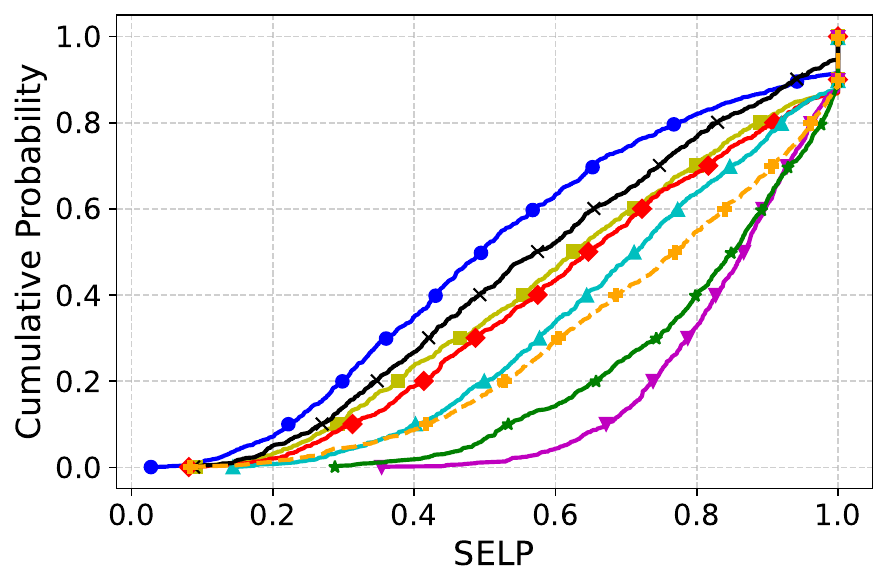}
        \caption{Network load $\lambda = 5$, with biased parameter $\tau = 0.5$ for the job selection.}
        \label{fig:cdf_5_dist_selp}
    \end{subfigure}
    \hfill
    \begin{subfigure}[b]{0.48\linewidth}
        \centering
        \includegraphics[width=\linewidth]{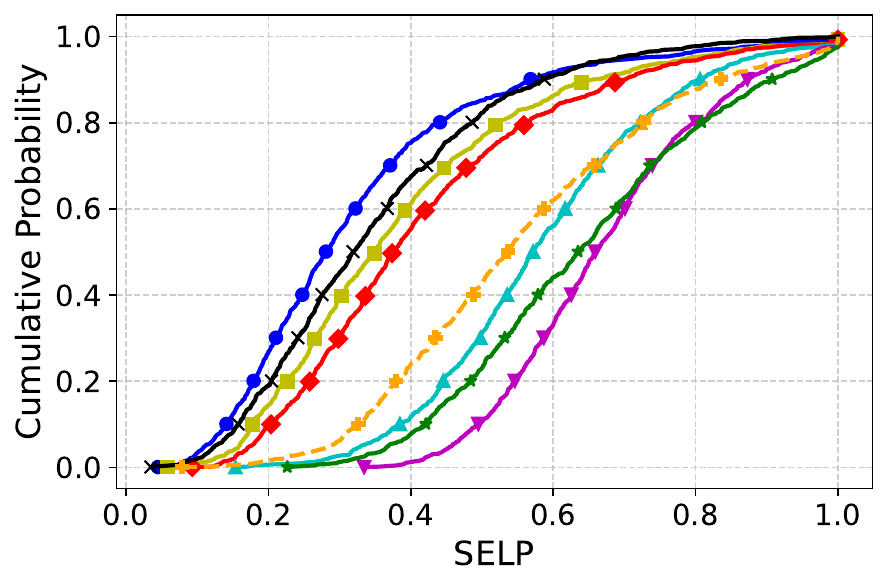}
        \caption{Network load $\lambda = 10$, with biased parameter $\tau = 0.5$ for the job selection.}
        \label{fig:cdf_10_dist_selp}
    \end{subfigure}
    \caption{The cumulative probability of SELP in each time slot among different schedulers.}
    \hrulefill
    \label{fig:selp}
\end{figure*}

\begin{figure*}
    \centering
    \begin{subfigure}[b]{0.48\linewidth}
        \centering
        \includegraphics[width=\linewidth]{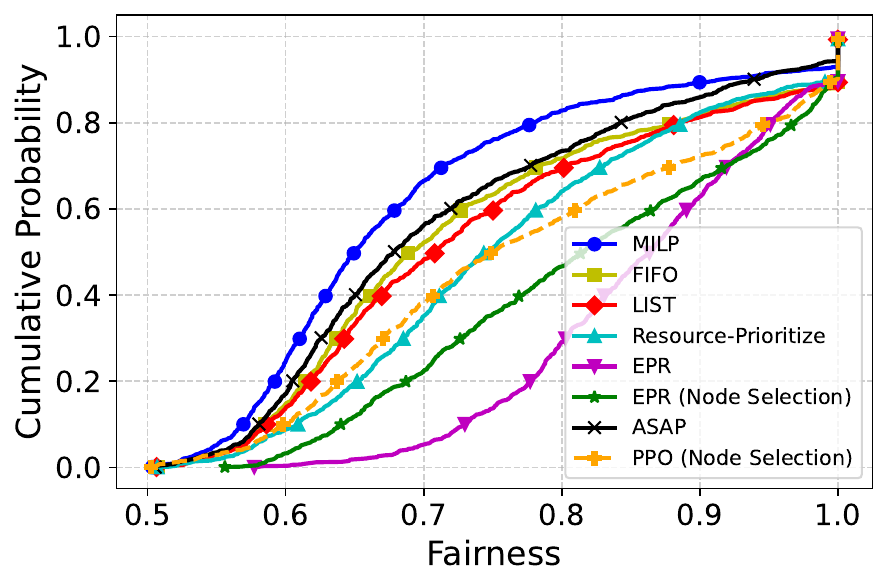}
        \caption{Network load $\lambda = 5$.}
        \label{fig:cdf_fairness}
    \end{subfigure}
    \hfill
    \begin{subfigure}[b]{0.48\linewidth}
        \centering
        \includegraphics[width=\linewidth]{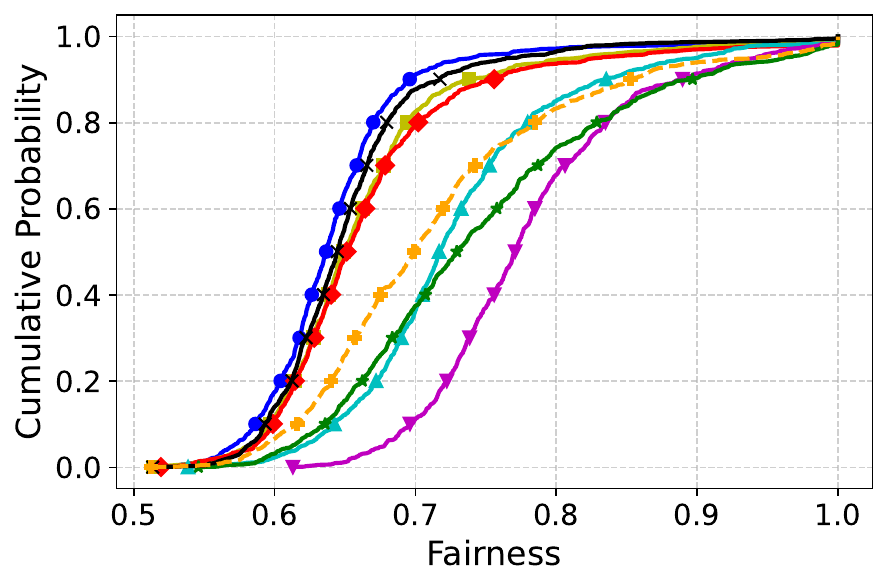}
        \caption{Network load $\lambda = 10$.}
        \label{fig:cdf_10_fairness}
    \end{subfigure}
    
    \begin{subfigure}[b]{0.48\linewidth}
        \centering
        \includegraphics[width=\linewidth]{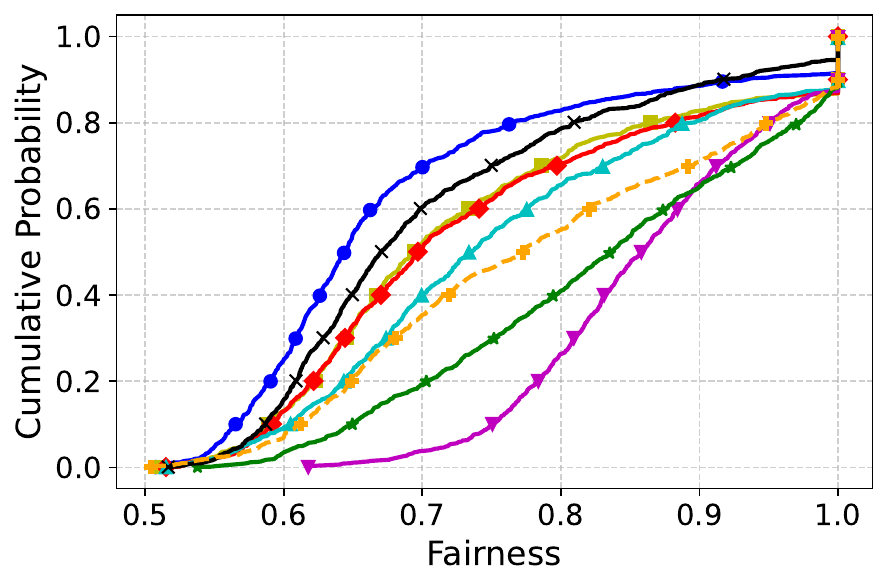}
        \caption{Network load $\lambda = 5$, with biased parameter $\tau = 0.5$ for the job selection.}
        \label{fig:cdf_5_dist_fairness}
    \end{subfigure}
    \hfill
    \begin{subfigure}[b]{0.48\linewidth}
        \centering
        \includegraphics[width=\linewidth]{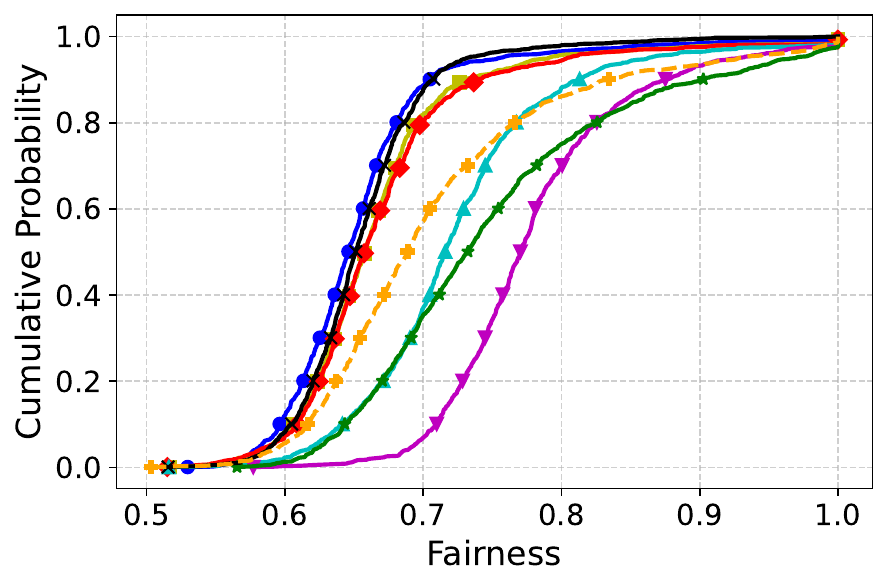}
        \caption{Network load $\lambda = 10$, with biased parameter $\tau = 0.5$ for the job selection.}
        \label{fig:cdf_10_dist_fairness}
    \end{subfigure}
    \caption{The cumulative probability of fairness in each time slot among different schedulers.}
    \hrulefill
    \label{fig:fairness}
\end{figure*}

The EPR scheduler achieves the highest SELP and fairness, followed by the EPR (node selection) scheduler, as shown in Fig.~\ref{fig:selp} and Fig.~\ref{fig:fairness}. This is because they prioritize jobs with shorter execution times and assign them to high-quality links, thereby avoiding prolonged execution that would result from scheduling on lower-quality links.

The Resource-Priority and PPO (node selection) exhibit the next best performance. In Table~\ref{tab:selp_1}, Table~\ref{tab:selp_2},  Table~\ref{tab:fair_1} and Table~\ref{tab:fair_2}, particularly under high network load, the Resource Priority scheduler achieves relatively higher SELP and fairness compared to the PPO (node selection) scheduler. This is due to its strategy of compacting as many jobs as possible while prioritizing combinations that consume fewer EPR pairs, resulting in relatively similar execution times and reduced waiting time for the remaining jobs.

It is important to note that the SELP and fairness metrics of the MILP and ASAP schedulers are lower than those of baseline methods such as LIST and FIFO. This behavior is expected, as both MILP and ASAP are formulated to minimize makespan exclusively, without explicitly accounting for execution latency or fairness considerations.

\begin{table*}
\centering
\footnotesize
\setlength{\tabcolsep}{6pt}
\begin{tabular}{l c c c c c c c c}
\hline
\textbf{Network settings}
& \multicolumn{2}{c}{\textbf{FIFO}}
& \multicolumn{2}{c}{\textbf{LIST}}
& \multicolumn{2}{c}{\textbf{Resource-Priority}}
& \multicolumn{2}{c}{\textbf{ASAP}} \\
\cline{2-9}
 & Mean & Std & Mean & Std & Mean & Std & Mean & Std \\
\hline

$\lambda = 5$
& 0.6331 & 0.2527
& 0.6609 & 0.2443
& 0.7201 & 0.2072
& 0.6097 & 0.2571 \\

$\lambda = 5$, bias $\tau = 0.5$
& 0.6285 & 0.2506
& 0.6472 & 0.2433
& 0.7009 & 0.2140
& 0.5865 & 0.2448 \\

$\lambda = 10$
& 0.3908 & 0.2011
& 0.4327 & 0.1998
& 0.6180 & 0.1605
& 0.3698 & 0.1880 \\

$\lambda = 10$, bias $\tau = 0.5$
& 0.3885 & 0.1937
& 0.4186 & 0.1929
& 0.5862 & 0.1634
& 0.3509 & 0.1742 \\
\hline
\end{tabular}
\caption{Mean and standard deviation of SELP under different network settings.}
\label{tab:selp_1}
\end{table*}

\begin{table*}
\centering
\footnotesize
\setlength{\tabcolsep}{6pt}
\begin{tabular}{l c c c c c c c c}
\hline
\textbf{Network settings}
& \multicolumn{2}{c}{\textbf{EPR}}
& \multicolumn{2}{c}{\textbf{EPR (Node Selection)}}
& \multicolumn{2}{c}{\textbf{MILP}}
& \multicolumn{2}{c}{\textbf{PPO (Node Selection)}} \\
\cline{2-9}
 & Mean & Std & Mean & Std & Mean & Std & Mean & Std \\
\hline

$\lambda = 5$
& 0.8476 & 0.1304
& 0.8026 & 0.1691
& 0.5461 & 0.2591
& 0.7254 & 0.2262 \\

$\lambda = 5$, bias $\tau = 0.5$
& 0.8443 & 0.1247
& 0.8089 & 0.1714
& 0.5323 & 0.2535
& 0.7331 & 0.2225 \\

$\lambda = 10$
& 0.6971 & 0.1423
& 0.6594 & 0.1783
& 0.3187 & 0.1925
& 0.5949 & 0.1899 \\

$\lambda = 10$, bias $\tau = 0.5$
& 0.6733 & 0.1434
& 0.6453 & 0.1796
& 0.3255 & 0.1858
& 0.5574 & 0.1945 \\
\hline
\end{tabular}
\caption{Mean and standard deviation of SELP under different network settings.}
\label{tab:selp_2}
\end{table*}

\begin{table*}
\centering
\footnotesize
\setlength{\tabcolsep}{6pt}
\begin{tabular}{l c c c c c c c c}
\hline
\textbf{Network settings}
& \multicolumn{2}{c}{\textbf{FIFO}}
& \multicolumn{2}{c}{\textbf{LIST}}
& \multicolumn{2}{c}{\textbf{Resource-Priority}}
& \multicolumn{2}{c}{\textbf{ASAP}} \\
\cline{2-9}
 & Mean & Std & Mean & Std & Mean & Std & Mean & Std \\
\hline

$\lambda = 5$
& 0.7348 & 0.1406
& 0.7432 & 0.1408
& 0.7661 & 0.1288
& 0.7195 & 0.1323 \\

$\lambda = 5$, bias $\tau = 0.5$
& 0.7370 & 0.1382
& 0.7410 & 0.1394
& 0.7618 & 0.1342
& 0.7107 & 0.1240 \\

$\lambda = 10$
& 0.6642 & 0.0757
& 0.6688 & 0.0805
& 0.7303 & 0.0798
& 0.6535 & 0.0631 \\

$\lambda = 10$, bias $\tau = 0.5$
& 0.6689 & 0.0712
& 0.6713 & 0.0730
& 0.7251 & 0.0759
& 0.6570 & 0.0516 \\
\hline
\end{tabular}
\caption{Mean and standard deviation of Fairness under different network settings.}
\label{tab:fair_1}
\end{table*}

\begin{table*}
\centering
\footnotesize
\setlength{\tabcolsep}{6pt}
\begin{tabular}{l c c c c c c c c}
\hline
\textbf{Network settings}
& \multicolumn{2}{c}{\textbf{EPR}}
& \multicolumn{2}{c}{\textbf{EPR (Node Selection)}}
& \multicolumn{2}{c}{\textbf{MILP}}
& \multicolumn{2}{c}{\textbf{PPO (Node Selection)}} \\
\cline{2-9}
 & Mean & Std & Mean & Std & Mean & Std & Mean & Std \\
\hline

$\lambda = 5$
& 0.8587 & 0.0952
& 0.8185 & 0.1302
& 0.6913 & 0.1281
& 0.7775 & 0.1455 \\

$\lambda = 5$, bias $\tau = 0.5$
& 0.8616 & 0.0891
& 0.8295 & 0.1269
& 0.6874 & 0.1304
& 0.7867 & 0.1422 \\

$\lambda = 10$
& 0.7815 & 0.0755
& 0.7478 & 0.1001
& 0.6452 & 0.0660
& 0.7169 & 0.0973 \\

$\lambda = 10$, bias $\tau = 0.5$
& 0.7816 & 0.0701
& 0.7514 & 0.0976
& 0.6545 & 0.0633
& 0.7095 & 0.0945 \\
\hline
\end{tabular}
\caption{Mean and standard deviation of Fairness under different network settings.}
\label{tab:fair_2}
\end{table*}

\section{Conclusion}
\label{sec:conlusion}

This paper presents an integrated workflow comprising a quantum compiler for compiling distributed quantum computation (DQC) jobs from quantum circuits, an Execution Manager for scheduling DQC jobs, and execution on the Qoala Simulator. As a key contribution, we propose multiple scheduling strategies within the Execution Manager to manage incoming DQC jobs, which are subsequently evaluated using various performance metrics.

The simulation results indicate that different schedulers exhibit distinct advantages. The EPR (node selection) and PPO (node selection) achieve the lowest makespan. The EPR scheduler achieves the best performance in system execution latency and fairness. The Resource-Priority, MILP, and ASAP schedulers achieve the highest QPU utilization. The MILP and ASAP schedulers perform second best in terms of makespan, while exhibiting the lowest performance in both system execution latency and fairness. The Resource-Priority scheduler shows moderate performance across makespan, system execution latency, and fairness. Unlike classical schedulers, PPO-based schedulers are driven by a reward function, offering greater flexibility in the optimization objective.

In future work, additional reward strategies could be investigated to further shape the behavior of PPO schedulers, with the aim of optimizing their intended objective (i.e., reducing the makespan, which is the primary goal of this paper). Moreover, the analysis of quantum network connectivity and topology, along with their effects on scheduling strategies, could be further investigated. In addition, with the advancement of quantum computers and their entanglement-based connections, DQC job schedulers on realistically connected quantum computer networks could be further evaluated and optimized.

\section*{Data Availability}
All data and code required to reproduce all plots shown herein are available at \url{https://doi.org/10.5281/zenodo.18802412}.

\section*{Acknowledgements}
This work was financially supported by the European Union's Horizon Europe research and innovation program under grant agreement No. 101102140 – QIA Phase 1, and by Research Ireland under Grant number \text{13/RC/2077\_P2}. Furthermore, this work benefited from the High Performance Computing facility at the University of Parma, Italy.

\bibliographystyle{unsrt}

\bibliography{references}

@inproceedings{ferrari2024execution,
  title={Execution management of distributed quantum computing jobs},
  author={Ferrari, Davide and Bandini, Michele and Amoretti, Michele},
  booktitle={2024 IEEE International Conference on Quantum Computing and Engineering (QCE)},
  volume={2},
  pages={150--154},
  year={2024},
  organization={IEEE}
}

@article{caleffi2024distributed,
  title={Distributed quantum computing: a survey},
  author={Caleffi, Marcello and Amoretti, Michele and Ferrari, Davide and Illiano, Jessica and Manzalini, Antonio and Cacciapuoti, Angela Sara},
  journal={Computer Networks},
  volume={254},
  pages={110672},
  year={2024},
  publisher={Elsevier}
}

@article{van2025qoala,
  title={Qoala: an application execution environment for quantum internet nodes},
  author={van der Vecht, Bart and Y{\"u}cel, Atak Talay and Jirovsk{\'a}, Hana and Wehner, Stephanie},
  journal={arXiv preprint arXiv:2502.17296},
  year={2025}
}

@inproceedings{chandra2024network,
  title={Network Operations Scheduling for Distributed Quantum Computing},
  author={Chandra, Nitish K and Kaur, Eneet and Seshadreesan, Kaushik P},
  booktitle={2024 IEEE 6th International Conference on Trust, Privacy and Security in Intelligent Systems, and Applications (TPS-ISA)},
  pages={506--515},
  year={2024},
  organization={IEEE}
}

@inproceedings{isard2009quincy,
  title={Quincy: fair scheduling for distributed computing clusters},
  author={Isard, Michael and Prabhakaran, Vijayan and Currey, Jon and Wieder, Udi and Talwar, Kunal and Goldberg, Andrew},
  booktitle={Proceedings of the ACM SIGOPS 22nd symposium on Operating systems principles},
  pages={261--276},
  year={2009}
}

@article{ferrari2023modular,
  title={A modular quantum compilation framework for distributed quantum computing},
  author={Ferrari, Davide and Carretta, Stefano and Amoretti, Michele},
  journal={IEEE Transactions on Quantum Engineering},
  volume={4},
  pages={1--13},
  year={2023},
  publisher={IEEE}
}

@inproceedings{xu2025remote,
  title={Remote Gate Scheduling in Distributed Quantum Computing},
  author={Xu, Xu and Liu, Yu and Mao, Yingling and Yang, Yuanyuan},
  booktitle={2025 IEEE 45th International Conference on Distributed Computing Systems (ICDCS)},
  pages={846--856},
  year={2025},
  organization={IEEE}
}

@article{johannes2006scheduling,
  title={Scheduling parallel jobs to minimize the makespan},
  author={Johannes, Berit},
  journal={Journal of Scheduling},
  volume={9},
  number={5},
  pages={433--452},
  year={2006},
  publisher={Springer}
}

@article{Brunner2014,
  title = {Bell nonlocality},
  author = {Brunner, Nicolas and Cavalcanti, Daniel and Pironio, Stefano and Scarani, Valerio and Wehner, Stephanie},
  journal = {Rev. Mod. Phys.},
  volume = {86},
  issue = {2},
  pages = {419--478},
  numpages = {60},
  year = {2014},
  month = {Apr},
  publisher = {American Physical Society},
  doi = {10.1103/RevModPhys.86.419},
  url = {https://link.aps.org/doi/10.1103/RevModPhys.86.419}
}

@ARTICLE{Amoretti2020,
  author={Amoretti, Michele and Carretta, Stefano},
  journal={IEEE Journal on Selected Areas in Communications}, 
  title={Entanglement Verification in Quantum Networks With Tampered Nodes}, 
  year={2020},
  volume={38},
  number={3},
  pages={598-604},
  doi={10.1109/JSAC.2020.2967955}
}

@article{adedoyin2018quantum,
  title={Quantum algorithm implementations for beginners},
  author={Adedoyin, Adetokunbo and Ambrosiano, John and Anisimov, Petr and Casper, William and Chennupati, Gopinath and Coffrin, Carleton and Djidjev, Hristo and Gunter, David and Karra, Satish and Lemons, Nathan and others},
  journal={arXiv preprint arXiv:1804.03719},
  year={2018}
}

@article{oslovich2025compilation,
  title={Compilation strategies for quantum network programs using Qoala},
  author={Oslovich, Samuel and van der Vecht, Bart and Wehner, Stephanie},
  journal={arXiv preprint arXiv:2505.06162},
  year={2025}
}

@inproceedings{dutton2008parallel,
  title={Parallel job scheduling with overhead: A benchmark study},
  author={Dutton, Richard A and Mao, Weizhen and Chen, Jie and Watson, William},
  booktitle={2008 international conference on networking, architecture, and storage},
  pages={326--333},
  year={2008},
  organization={IEEE}
}

@inproceedings{sgall2014multiprocessor,
  title={Multiprocessor jobs, preemptive schedules, and one-competitive online algorithms},
  author={Sgall, Ji{\v{r}}{\'\i} and Woeginger, Gerhard J},
  booktitle={International Workshop on Approximation and Online Algorithms},
  pages={236--247},
  year={2014},
  organization={Springer}
}

@article{schulman2017proximal,
  title={Proximal policy optimization algorithms},
  author={Schulman, John and Wolski, Filip and Dhariwal, Prafulla and Radford, Alec and Klimov, Oleg},
  journal={arXiv preprint arXiv:1707.06347},
  year={2017}
}

@article{cuomo2023optimized,
  title={Optimized compiler for distributed quantum computing},
  author={Cuomo, Daniele and Caleffi, Marcello and Krsulich, Kevin and Tramonto, Filippo and Agliardi, Gabriele and Prati, Enrico and Cacciapuoti, Angela Sara},
  journal={ACM Transactions on Quantum Computing},
  volume={4},
  number={2},
  pages={1--29},
  year={2023},
  publisher={ACM New York, NY}
}

@article{diadamo2021distributed,
  title={Distributed quantum computing and network control for accelerated vqe},
  author={DiAdamo, Stephen and Ghibaudi, Marco and Cruise, James},
  journal={IEEE Transactions on Quantum Engineering},
  volume={2},
  pages={1--21},
  year={2021},
  publisher={IEEE}
}

@inproceedings{orenstein2024qgroup,
  title={Qgroup: Parallel quantum job scheduling using dynamic programming},
  author={Orenstein, Aaron and Chaudhary, Vipin},
  booktitle={2024 IEEE International conference on quantum computing and engineering (QCE)},
  volume={1},
  pages={990--999},
  year={2024},
  organization={IEEE}
}

@inproceedings{parekh2021quantum,
  title={Quantum algorithms and simulation for parallel and distributed quantum computing},
  author={Parekh, Rhea and Ricciardi, Andrea and Darwish, Ahmed and DiAdamo, Stephen},
  booktitle={2021 IEEE/ACM Second International Workshop on Quantum Computing Software (QCS)},
  pages={9--19},
  year={2021},
  organization={IEEE}
}

@article{Krutyanskiy2023Entanglement,
  title = {Entanglement of Trapped-Ion Qubits Separated by 230 Meters},
  author = {Krutyanskiy, V. and Galli, M. and Krcmarsky, V. and Baier, S. and Fioretto, D. A. and Pu, Y. and Mazloom, A. and Sekatski, P. and Canteri, M. and Teller, M. and Schupp, J. and Bate, J. and Meraner, M. and Sangouard, N. and Lanyon, B. P. and Northup, T. E.},
  journal = {Phys. Rev. Lett.},
  volume = {130},
  issue = {5},
  pages = {050803},
  numpages = {7},
  year = {2023},
  month = {Feb},
  publisher = {American Physical Society},
  doi = {10.1103/PhysRevLett.130.050803},
  url = {https://link.aps.org/doi/10.1103/PhysRevLett.130.050803}
}

@article{turkakin2021comparison,
  author={T{\"u}rkak{\i}n, Osman H{\"u}rol and Arditi, David and Manisal{\i}, Ekrem},
  title={Comparison of heuristic priority rules in the solution of the resource-constrained project scheduling problem},
  journal={Sustainability},
  volume={13},
  number={17},
  pages={9956},
  year={2021},
  publisher={MDPI}
}

@article{topcuoglu2002performance,
  author={Topcuoglu, Haluk and Hariri, Salim and Wu, Min-You},
  title={Performance-effective and low-complexity task scheduling for heterogeneous computing},
  journal={IEEE transactions on parallel and distributed systems},
  volume={13},
  number={3},
  pages={260--274},
  year={2002},
  publisher={IEEE}
}

@article{tesch2020polyhedral,
  author={Tesch, Alexander},
  title={A polyhedral study of event-based models for the resource-constrained project scheduling problem},
  journal={Journal of Scheduling},
  volume={23},
  number={2},
  pages={233--251},
  year={2020},
  publisher={Springer}
}

@inproceedings{feitelson1995parallel,
  author={Feitelson, Dror G and Rudolph, Larry},
  title={Parallel job scheduling: Issues and approaches},
  booktitle={Workshop on Job Scheduling Strategies for Parallel Processing},
  pages={1--18},
  year={1995},
  organization={Springer}
}

@inproceedings{lifka1995anl,
  author={Lifka, David A},
  title={The anl/ibm sp scheduling system},
  booktitle={Workshop on Job Scheduling Strategies for Parallel Processing},
  pages={295--303},
  year={1995},
  organization={Springer}
}

@inproceedings{yu2022surprising,
  author={Yu, Chao and Velu, Akash and Vinitsky, Eugene and Gao, Jiaxuan and Wang, Yu and Bayen, Alexandre and Wu, Yi},
  title={The Surprising Effectiveness of PPO in Cooperative Multi-Agent Games},
  booktitle={Advances in Neural Information Processing Systems},
  volume={35},
  pages={24611--24624},
  year={2022}
}

@inproceedings{kool2019stochastic,
  author={Kool, Wouter and Van Hoof, Herke and Welling, Max},
  title={Stochastic Beams and Where to Find Them: The Gumbel-Top-k Trick for Sampling Sequences Without Replacement},
  booktitle={ICML},
  year={2019}
}

@ARTICLE{bandini2024optimized,
  author={Bandini, Michele and Ferrari, Davide and Carretta, Stefano and Amoretti, Michele},
  journal={IEEE Access}, 
  title={Optimized Compilation for Distributed Quantum Computing}, 
  year={2026},
  volume={},
  number={},
  pages={1-1},
  doi={10.1109/ACCESS.2026.3706203}
}

\pagebreak
\appendix
\section{Specific DQC job types}
\label{sec:apendix_DQC_job_types}
\subsection{GHZ State}

The Greenberger–Horne–Zeilinger (GHZ) state is an entangled multi-qubit state of the form:

\begin{equation}
|\mathrm{GHZ}_n\rangle = \frac{|0\rangle^{\otimes n} + |1\rangle^{\otimes n}}{\sqrt{2}}.
\end{equation}

It can be prepared on $n$ qubits by first applying a Hadamard gate to the first qubit, followed by a sequence of CNOT gates connecting the first qubit to all subsequent qubits:

\begin{equation}
|\mathrm{GHZ}_n\rangle = \text{CNOT}_{1,2} \cdots \text{CNOT}_{1,n} \, H_1 \, |0\rangle^{\otimes n}.
\end{equation}

A $5$-qubit GHZ circuit using CNOT gates is illustrated schematically in Fig.~\ref{fig:ghz_cx}.

\begin{figure}[!ht]
\centering
\begin{adjustbox}{width=0.5\linewidth}
\begin{quantikz}[row sep=0.5cm, column sep=0.5cm]
\lstick{\ket{q_0}} & \gate{H} & \ctrl{1} & \ctrl{2} & \ctrl{3} & \ctrl{4} & \qw \\
\lstick{\ket{q_1}} & \qw      & \targ{} & \qw      & \qw      & \qw      & \qw \\
\lstick{\ket{q_2}} & \qw      & \qw      & \targ{} & \qw      & \qw      & \qw \\
\lstick{\ket{q_3}} & \qw      & \qw      & \qw      & \targ{} & \qw      & \qw \\
\lstick{\ket{q_4}} & \qw      & \qw      & \qw      & \qw      & \targ{} & \qw
\end{quantikz}
\end{adjustbox}
\caption{$5$-qubit GHZ state preparation. The first qubit is put into superposition via a Hadamard gate, and subsequent qubits are entangled via controlled-NOT operations.}
\hrulefill
\label{fig:ghz_cx}
\end{figure}

\subsection{Graph State}

A ring graph state associated with a graph $G (V,E)$ is defined as

\begin{equation}
    |G\rangle 
    = 
    \left( \prod_{(i,j)\in E} \mathrm{CZ}_{i,j} \right)
    |+\rangle^{\otimes n},
    \label{eq:graph_state_def}
\end{equation}

where $n = |V|$ is the number of vertices. Each qubit is initialized in the $|+\rangle$ state via Hadamard gates, after which entangling controlled-Z operations are applied for every edge $(i,j) \in E$. The quantum circuit is shown in Fig.~\ref{fig:graph_state}.

\begin{figure}[!ht]
\centering
\begin{adjustbox}{width=0.5\linewidth}
\begin{quantikz}[row sep=0.5cm, column sep=0.5cm]
\lstick{\ket{q_0}} & \gate{H} & \ctrl{1} & \qw & \qw & \qw      & \ctrl{4} & \qw \\
\lstick{\ket{q_1}} & \gate{H} & \gate{Z} & \ctrl{1} & \qw & \qw      & \qw      & \qw \\
\lstick{\ket{q_2}} & \gate{H} & \qw      & \gate{Z} & \ctrl{1} & \qw    & \qw      & \qw \\
\lstick{\ket{q_3}} & \gate{H} & \qw      & \qw      & \gate{Z} & \qw    & \qw      & \qw \\
\lstick{\ket{q_4}} & \gate{H} & \qw      & \qw      & \qw      & \qw    & \gate{Z} & \qw
\end{quantikz}
\end{adjustbox}
\caption{$5$-qubit graph state preparation. Each qubit is initialized in the $|+\rangle$ state via a Hadamard gate, and controlled-Z (CZ) gates is introduced according to the edges of the graph.}
\hrulefill
\label{fig:graph_state}
\end{figure}

\subsection{Quantum Approximate Optimization Algorithm (QAOA)}

This algorithm is designed to provide approximate solutions to combinatorial optimization problems~\cite{adedoyin2018quantum}. It includes the cost Hamiltonian $\hat{H}_c = \sum_{\alpha} C_\alpha$, where $C_\alpha$ is the quantum clause Hamiltonian, and a mixer Hamiltonian $\hat{H}_b = \sum_{j} X_j$, where $X_j$ denotes the Pauli-$X$ operator acting on qubit $j$.

Given a number of rounds $r \ge 1$ and parameter sets $\{\beta_k\}_{k=1}^{r}$ and $\{\gamma_k\}_{k=1}^{r}$, the algorithm starts in the uniform superposition 
$|+\rangle^{\otimes n}$ and alternates the application of the cost and mixer unitaries:
\begin{align}
U_C(\gamma_k) &=  e^{-i\gamma_k \hat{H}_c}, \\
U_B(\beta_k) &=  e^{-i\beta_k \hat{H}_b}.
\end{align}

This produces the variational state $|\boldsymbol{\beta}, \boldsymbol{\gamma}\rangle$, from which candidate solutions are sampled. The quality of these solutions depends on the chosen round $r$ and the parameters $\boldsymbol{\beta}, \boldsymbol{\gamma}$.

For the MaxCut problem, the cost Hamiltonian $\hat{H}_c = \sum_{(i,j) \in E} \frac{1}{2} (I - Z_i Z_j)$.  An illustrative QAOA circuit for a $5$-qubit system on the simple ring graph with edges $(0,1)$, $(1,2)$, $(2,3)$, $(3,4)$, and $(0,4)$ is shown in Fig.~\ref{fig:qaoa_circuit}. Specifically, the rotation gates applied in the quantum circuit are defined as:

\begin{flalign}
R_c(\gamma) &= \begin{pmatrix} 1 & 0 \\ 0 & e^{-i\gamma} \end{pmatrix}, &\\
R_b(\theta, \phi, 2\beta) &= \begin{pmatrix} \cos (\theta / 2) & -e^{2i \beta} \sin (\theta / 2) \\ e^{i \phi} \sin (\theta / 2) & e^{i(2\beta+\phi)} \cos (\theta / 2) \end{pmatrix}. &
\end{flalign}

\begin{figure}[!ht]
\centering
\begin{adjustbox}{width=1\linewidth}
\begin{quantikz}[row sep=0.5cm, column sep=0.5cm]
\lstick{\ket{q_0}} & \gate{H} & \ctrl{1} & \qw & \ctrl{1} & \qw & \qw & \qw & \qw & \qw & \qw & \qw & \qw & \qw & \gate{R_b(2\beta)} & \qw \\
\lstick{\ket{q_1}} & \gate{H} & \targ{} & \gate{R_c(\gamma)} & \targ{} & \ctrl{1} & \qw & \ctrl{1} & \qw & \qw & \qw & \qw & \qw & \qw & \gate{R_b(2\beta)} & \qw \\
\lstick{\ket{q_2}} & \gate{H} & \qw & \qw & \qw & \targ{} & \gate{R_c(\gamma)} & \targ{} & \ctrl{1} & \qw & \ctrl{1} & \qw & \qw & \qw & \gate{R_b(2\beta)} & \qw \\
\lstick{\ket{q_3}} & \gate{H} & \qw & \qw  & \qw & \qw  & \qw & \qw  & \targ{} & \gate{R_c(\gamma)} & \targ{} & \ctrl{1} & \qw & \ctrl{1} & \gate{R_b(2\beta)} & \qw \\
\lstick{\ket{q_4}} & \gate{H} & \qw & \qw  & \qw & \qw & \qw & \qw & \qw & \qw & \qw & \targ{} & \gate{R_c(\gamma)} & \targ{} & \gate{R_b(2\beta)} & \qw
\end{quantikz}
\end{adjustbox}
\caption{$5$-qubit QAOA circuit with one repetition. The cost unitary $U_C(\gamma)$ is implemented via two controlled-NOT and one $R_c(\gamma)$ gates, which is a rotation that depends on the parity of the two qubits.The mixer unitary $U_B(\beta)$ is applied as $R_b(2\beta)$ rotations on all qubits.}
\hrulefill
\label{fig:qaoa_circuit}
\end{figure}

\subsection{Quantum Fourier Transform (QFT)}

The Quantum Fourier Transform is defined as:

\begin{equation}
\label{eq:qft}
\mathrm{QFT}\ket{x} = \frac{1}{\sqrt{2^m}} \sum_{y=0}^{2^m-1} e^{2\pi i xy / 2^m} \ket{y},
\end{equation}
where the amplitudes associated with the $\ket{y}$ vectors are the discrete Fourier transforms of the amplitudes of the $\ket{x}$ vectors.

The QFT is implemented recursively, starting with a Hadamard gate applied to the first qubit to create a superposition. Each subsequent qubit is then rotated conditionally using controlled rotations $R_{m}=\left(\begin{array}{cc}1 & 0 \\ 0 & e^{2 \pi i / 2^{m}}\end{array}\right)$ on the target qubit. A schematic $5$-qubit QFT circuit is shown in Fig.~\ref{fig:qft_circuit}.

\begin{figure}[!ht]
\centering
\begin{adjustbox}{width=0.98\linewidth}
\begin{quantikz}[row sep=0.5cm, column sep=0.5cm]
\lstick{\ket{q_4}} & \gate{H} & \ctrl{1} & \ctrl{2} & \ctrl{3} & \ctrl{4} & \qw & \qw & \qw & \qw & \qw & \qw & \qw & \qw & \qw & \qw & \qw \\
\lstick{\ket{q_3}} & \qw & \gate{R_{2}} & \qw & \qw & \qw & \gate{H} & \ctrl{1} & \ctrl{2} & \ctrl{3} & \qw & \qw & \qw & \qw & \qw & \qw & \qw \\
\lstick{\ket{q_2}} & \qw & \qw & \gate{R_{3}} & \qw & \qw & \qw & \gate{R_{2}} & \qw & \qw & \gate{H} & \ctrl{1} & \ctrl{2} & \qw & \qw & \qw & \qw \\
\lstick{\ket{q_1}} & \qw & \qw & \qw & \gate{R_{4}} & \qw & \qw & \qw & \gate{R_{3}} & \qw & \qw & \gate{R_{2}} & \qw & \gate{H} & \ctrl{1} & \qw & \qw \\
\lstick{\ket{q_0}} & \qw & \qw & \qw & \qw & \gate{R_{5}} & \qw & \qw & \qw & \gate{R_{4}} & \qw & \qw & \gate{R_{3}} & \qw & \gate{R_{2}} & \gate{H} & \qw
\end{quantikz}
\end{adjustbox}
\caption{$5$-qubit QFT circuit with controlled-phase rotations $R_k$. Hadamard gates initialize superpositions, and controlled-phase gates implement the Fourier transform phases.}
\hrulefill
\label{fig:qft_circuit}
\end{figure}

\subsection{Variational Quantum Eigen-solver (VQE)}

This is a hybrid quantum-classical algorithm designed to approximate the ground state energy of a given Hamiltonian $\hat{H}$, as described in Ref.~\cite{adedoyin2018quantum}. A variational state $\lvert \psi(\theta_i) \rangle$ is prepared with parameters $\theta_i$, chosen such that the number of parameters scales linearly with the system size. The expectation value of the Hamiltonian is calculated on a quantum computer as
\begin{equation}
    E = \frac{\langle \psi(\theta) \lvert \hat{H} \rvert \psi(\theta) \rangle}{\langle \psi(\theta) \vert \psi(\theta) \rangle}.
\end{equation}
The parameters $\theta_i$ are updated using a classical nonlinear optimizer. In this paper, EfficientSU2 is used, for which each qubit is first rotated using parameterized $R_Y$ and $R_Z$ gates with parameters randomly initialized in $[0, 2\pi]$.  Entanglement is applied linearly, such that controlled-NOT gates connect each qubit to its nearest neighbor sequentially. A schematic representation of a single-layer $5$-qubit VQE is shown in Fig.~\ref{fig:vqe_circuit}.

\begin{figure}[!ht]
\centering
\begin{adjustbox}{width=0.5\linewidth}
\begin{quantikz}[row sep=0.5cm, column sep=0.4cm]
\lstick{\ket{q_0}} & \gate{R_Y(\theta_1)} & \gate{R_Z(\theta_2)} & \ctrl{1} & \qw & \qw & \qw & \qw \\
\lstick{\ket{q_1}} & \gate{R_Y(\theta_3)} & \gate{R_Z(\theta_4)} & \targ{} & \ctrl{1} & \qw & \qw & \qw \\
\lstick{\ket{q_2}} & \gate{R_Y(\theta_5)} & \gate{R_Z(\theta_6)} & \qw & \targ{} & \ctrl{1} & \qw & \qw\\
\lstick{\ket{q_3}} & \gate{R_Y(\theta_7)} & \gate{R_Z(\theta_8)} & \qw & \qw & \targ{} & \ctrl{1} & \qw \\
\lstick{\ket{q_4}} & \gate{R_Y(\theta_9)} & \gate{R_Z(\theta_{10})} & \qw & \qw & \qw & \targ{} & \qw
\end{quantikz}
\end{adjustbox}
\caption{$5$-qubit VQE circuit using linear nearest-neighbor entanglement with one repetition. Each qubit undergoes parameterized $R_Y$ and $R_Z$ rotations, followed by controlled-NOT gates between nearest neighbors. The rotation parameters $\theta_i$ are randomly initialized in $[0, 2\pi]$.}
\hrulefill
\label{fig:vqe_circuit}
\end{figure}

\end{document}